\documentclass[aps,prc,twocolumn,amssymb,amsmath,superscriptaddress,floatfix,showpacs,nofootinbib]{revtex4}
\usepackage[dvipdfm]{graphicx}
\usepackage{multirow}
\usepackage{longtable}

\begin{document}
\title{Charmonium mass in hot and dense hadronic matter}
\author{Kenji Morita}
\email{kmorita@yukawa.kyoto-u.ac.jp}
\affiliation{GSI, Helmholzzentrum f\"{u}r Schwerionenforschung,
Planckstr. 1, D-64291 Darmstadt, Germany}
\affiliation{Yukawa Institute for Theoretical
Physics, Kyoto University, Kyoto 606-8502, Japan}

\author{Su Houng Lee}
\email{suhoung@yonsei.ac.kr}
\affiliation{Institute of Physics and Applied Physics, Yonsei
University, Seoul 120-749, Korea}

\date{\today}
\begin{abstract}
We investigate mass shifts of charmonia driven by
 change of the gluon condensate below but near transition
 temperatures at finite baryonic chemical potential. Extending previous prescription on the
 relation between gluon
 condensates and thermodynamic quantities, we model the gluon
 condensates of hadronic matter at finite temperature and baryonic chemical
 potential such that the scalar gluon condensate fits with the latest lattice
 QCD data. By making use of the QCD sum rule and the second order Stark
 effect, we find that the smoother transition in the full QCD can lead to
 moderate mass shifts of charmonia even below the transition
 temperature. We also find larger mass shift
 at fixed temperature as chemical potential increases.
 Existing data on charmonium-charmonium ratio is found to be consistent
 with the statistical hadronization scenario including the obtained mass shift.
\end{abstract}

\maketitle

\section{Introduction}
\label{sec:intro}

Properties of heavy quarkonia in medium have been extensively studied
since it was pointed out that suppression of $J/\psi$ by Debye screening
could be a signature of creation of the deconfined matter in
relativistic heavy ion collisions \cite{Matsui_PLB178}. It should be
noted that a mass shift of a charmonium state in hot hadronic
environment could be a precursor phenomenon of the transition caused by
a decrease in the string tension \cite{hashimoto86}. These early
expectations are based on
an intuitive picture on a quarkonium, a heavy quark and its antiquark
bound by a confining potential, which successfully describes the
properties in vacuum \cite{eichten78:_charm}.
While the lattice QCD provides a first principle approach to the
problem, it still lacks the necessary resolution needed for discriminating possible changes in the quarkonium
spectral function at finite temperature, especially near the critical temperature where an abrupt change could take place. The maximum entropy method for this problem can at best only
tell us about the (non-)existence of the lowest peak in the spectral
function \cite{Asakawa_PRL92,Datta_PRD69,umeda05,jakovac07,aarts07}.
Therefore, to assess the medium modification one needs a complementary
framework such as the potential model which utilizes a quark-antiquark
potential extracted from lattice calculation \cite{mocsy:_poten}.
In the meantime, we have proposed an approach utilizing the QCD sum rule
and the second order Stark effect which allows us to relate the
temperature dependent gluon condensates as the primary inputs from
lattice QCD to the spectral changes of heavy quarkonia
\cite{morita_jpsiprl,morita_jpsifull,song09,lee_morita_stark,morita_borel}.

So far our studies relied on the gluon condensates extracted from the
trace anomaly of pure SU(3) case \cite{Boyd_NPB469}. In pure gauge theory with
$N_c \geq 3$, there is a first order deconfinement transition at $T=T_c$
thus the trace anomaly shows abrupt change across $T_c$ \cite{panero09:_therm_qcd_n}, which leads to
the similar behavior of the spectral property of the quarkonia.

In order to compare results with experimental data, we need more realistic estimates of the condensates
based on full QCD lattice calculations including dynamical light quarks.
Recently the calculation of the equation of state
has been carried out with physical quark masses at vanishing chemical
potential \cite{borsanyi10:_qcd}. Due to
the crossover nature of the transition~\cite{aoki06}, the critical
temperature is no longer a well-defined quantity. The pseudocritical
temperature defined by a peak or an inflection point depends on
observables  \cite{borsanyi10:_in_t_qcd} and is found to range from
147 MeV (chiral susceptibility) to 165 MeV (strange
quark number susceptibility).

As for the scalar gluon
condensate, which is the gluonic part of the trace anomaly, it is
expected to have the same bulk property but with smoother change near
the pseudocritical temperature $T_{pc}$.
In fact, the magnitude for the change of the scalar condensate in a pure
gauge theory at the transition region has been found to be almost the same as in the
full QCD case \cite{morita_jpsifull,cheng08} when the temperatures are
normalized by $T_c$ and $T_{pc}$, despite the difference in the (pseudo)critical temperatures;
$T_c\simeq 265$ MeV in the pure SU(3) \cite{Boyd_NPB469,panero09:_therm_qcd_n} and
$T_{pc}\simeq 190-200$ MeV in full (2+1) flavor QCD with $m_\pi\simeq
220$ MeV \cite{cheng08}. The crossover nature of the
transition has led to the smoother temperature dependence of the gluon
condensate near $T_{pc}$ \cite{bazavov09:_equat_qcd}. For the twist-2
gluon condensate,  which is not a dominant but non-negligible
contribution to the sum rule \cite{Hatsuda93,morita_jpsifull},
no lattice data at physical quark masses is available yet.

In this paper, we first extract the gluon condensates in full QCD by making
use of the resonance gas model. This prescription enables us to study finite
baryonic chemical potential case also. Using these gluon condensates,
we investigate the change of spectral properties of charmonia in hot
hadronic matter at various values of temperature and baryonic chemical
potential along the freeze-out line which is deduced from statistical model
analyses \cite{cleymans06:_compar,andronic06:_hadron} and is expected to be close to the
hadronization points. Although the charmonium production mechanism in
relativistic heavy ion collisions has not been understood well, some
experimental data seem to indicate that the statistical production of charmonia
could be possible \cite{andronic09:_statis}. Despite the complexity of the collision
processes, this scenario considerably simplifies the
charmonium-charmonium particle number ratio. We examine possible
influences of the spectral modification on this quantity.

In the next section, we present a resonance gas model for the gluon
condensates of hot and dense hadronic matter. In
Sec~\ref{sec:massshift}, we
report results of spectral changes of charmonia. The experimental
implication will be discussed in Sec.~\ref{sec:imp_exp}. Section
\ref{sec:summary} is devoted to a summary.

\section{Resonance gas model for the gluon condensates}

We start with two quantities $M_0$ and $M_2$ characterizing the
temperature dependence of the thermal expectation value of gluonic
operators
\begin{align}
 \left\langle \frac{\beta(g)}{2g}G^a_{\mu\nu}G^{a\mu\nu} \right\rangle_T &= M_0(T)\\
 \left\langle -\mathcal{ST} G^{a}_{\alpha\mu}G_\beta^{a\mu}
 \right\rangle_T &= \left(u_\alpha u_\beta - \frac{1}{4}g_{\alpha\beta}\right)M_2(T). \label{def-m}
\end{align}
Here, the symbol $\mathcal{ST}$ denotes the traceless and symmetric part
of the operator. We assign to $M_0(T)$ only the temperature dependent part of
the expectation value; it has in general a temperature independent part
which is nothing but the gluon condensate in the vacuum.

The above equations immediately relate $M_0$ and $M_2$
to the thermodynamic quantities via the
energy-momentum tensor in thermal equilibrium; namely,
$M_0(T)=\varepsilon-3p$ and $M_2(T)=\varepsilon+p$ with $\varepsilon$
and $p$ being the energy density and the pressure respectively in the
case of pure gluonic system \cite{lee_morita_stark}. In the presence of
fermions, however, it is not straightforward to relate $M_0$ and $M_2$
to the thermodynamic quantities since the energy-momentum tensor has
fermionic part. The trace anomaly receives contributions from
massive fermions as
\begin{equation}
 \langle T_\mu^\mu \rangle = \left\langle \frac{\beta(g)}{2g}
			      G^a_{\mu\nu}G^{a\mu\nu} \right\rangle +
 \sum_{i}m_i \langle \bar{q}_iq_i \rangle.\label{eq:e-mtensor}
\end{equation}
$M_0$ becomes $\varepsilon-3p$ only when the second term is
negligible, otherwise, the fermionic part has to be explicitly
subtracted out from $\varepsilon-3p$ before identifying it to $M_0$. In
lattice QCD calculations, contributions from each term in
Eq.~\eqref{eq:e-mtensor} have been
estimated \cite{cheng08,bazavov09:_equat_qcd}\footnote{In
Ref.~\cite{borsanyi10:_qcd}, however, somewhat different scheme is used
to calculate the trace anomaly.}.
Similarly, $M_2$ can be related to the off-diagonal part of the
energy-momentum tensor after the fermionic part is subtracted out.  Such
data are not yet available. Therefore we need a scheme to subtract the
fermionic contribution from the total of the energy-momentum tensor.

Going back to the original definition given in Eq.~\eqref{def-m}, one
can relate the nucleon expectation values to $M_0$ and $M_2$ within the linear
density approximation \cite{Klingl_PRL82}
\begin{align}
 M_0^{\text{n.m.}}&= \rho m_N^0,\label{eq:m0_nm}\\
 M_2^{\text{n.m.}}&= \rho A_G m_N,\label{eq:m2_nm}
\end{align}
where $\rho$, $m_N^0$, $A_G$ and $m_N$ are the density of nucleus,
the nucleon mass in the chiral limit, the second moment of gluon
distribution function of the nucleon, and the nucleon mass,
respectively. One sees in
Eq.~\eqref{eq:m0_nm} that the chiral limit is taken for the nucleon mass, which
corresponds to removing the fermionic term in the trace anomaly
[Eq.~\eqref{eq:e-mtensor}]. In Eq.~\eqref{eq:m2_nm}, $A_G$ plays a
similar role. Since these equations are expressed in terms of the mass of
the particle consisting of the medium and its number density, one can extend
them to genuine hadronic matter as
\begin{align}
 M_0^{\text{had}}&= \sum_{i=\text{hadrons}} \rho_i m_i^0\label{eq:m0_model}\\
 M_2^{\text{had}}&= \sum_{i=\text{hadrons}} \rho_i m_i A_G^i.\label{eq:m2_model}
\end{align}
The quantities with subscript $i$ denote hadronic counterparts for
the nucleon values appearing in Eqs.~\eqref{eq:m0_nm} and
\eqref{eq:m2_nm}.
As a simple model, we use a hadron resonance gas including all
hadrons for which the quantum numbers are known as given
in the Particle Data Group \cite{pdg2010}.
The number density $\rho_i$ is now generalized to a
hadron gas and is calculated as a function of $T$ and
$\mu_B$ as
\begin{equation}
 \rho_i = \frac{d_i}{2\pi^2}\intop_0^{\infty}
  \frac{p^2dp}{\exp[(\sqrt{p^2+m_i^2}-\mu_i)/T]\pm 1}
\end{equation}
where the sign is $+$ for fermions and $-$ for bosons and $d_i$ is the
degree of freedom of the $i$-th hadron. We take into account the baryon
number conservation and strangeness conservation and neglect isospin
chemical potential for simplicity. The strangeness chemical potential
$\mu_s$
is determined from the neutrality condition $\sum_{i}\rho_i S_i=0$ \cite{koch83:_stran}.

In Eq.~\eqref{eq:m0_model}, the masses of hadrons in the
chiral limit $m_0^i$ are needed. Note that
strange quark contribution from Eq.~\eqref{eq:e-mtensor} and its
off-diagonal counterpart also have to be subtracted. Thus we will work
within the flavor SU(3) symmetric limit.
At present, we cannot know all of hadron masses in the $m_u=m_d=m_s=0$ limit,
especially those of the highly excited states. Therefore for the masses
in the three-flavor chiral limit, we use different masses only for the Goldstone
bosons, ground state octet and decuplet baryons,
and keep the masses of other hadrons the same as their vacuum
values. Detailed lattice studies on hadron masses, as done in
Ref.~\cite{Durr08:_science} with physical quark masses, will be helpful
for more accurate treatment.
Specifically, we first put $m_\pi^0=m_K^0=0$, and $m_N^0=750$ MeV from
heavy baryon chiral perturbation theory \cite{borasoy96:_baryon}; these
are the most important inputs needed for
the masses in the chiral limit as the contributions to the thermodynamic
quantities are dominated by these hadrons, especially by the
Nambu-Goldstone bosons.  For the vector and axial
vector mesons, we assume $m_\rho^0=m_\rho$ and $m_{a_1}^0=m_{a_1}$.
We also assume $m_\Delta^0=m_\Delta$.
Furthermore, we also put $m_{f_0}^0 = m^0_{\sigma}$,
$m_\phi^0=m_\omega^0 = m_{K^*}^0=m_\rho^0$,
$m_\Lambda^0=m_\Xi^0=m_\Sigma^0=m_N^0$,
$m_{\Sigma^*}^0=m_{\Xi^*}^0=m_{\Omega}^0=m_\Delta^0$
according to the flavor SU(3) symmetry.

The second moment of the gluon distribution function is set to $A_G^i(8m_c^2) = 0.9$
for all the hadrons. Generally it can differ among hadrons, but it can
be shown that $A_G^{\pi}$ differs little from this value at such a high energy scale
where the parametrization of the gluon distribution functions
\cite{gluck92:_parton,gluck99:_pionic} are relatively well known.

\begin{figure}[!t]
 \includegraphics[width=3.375in]{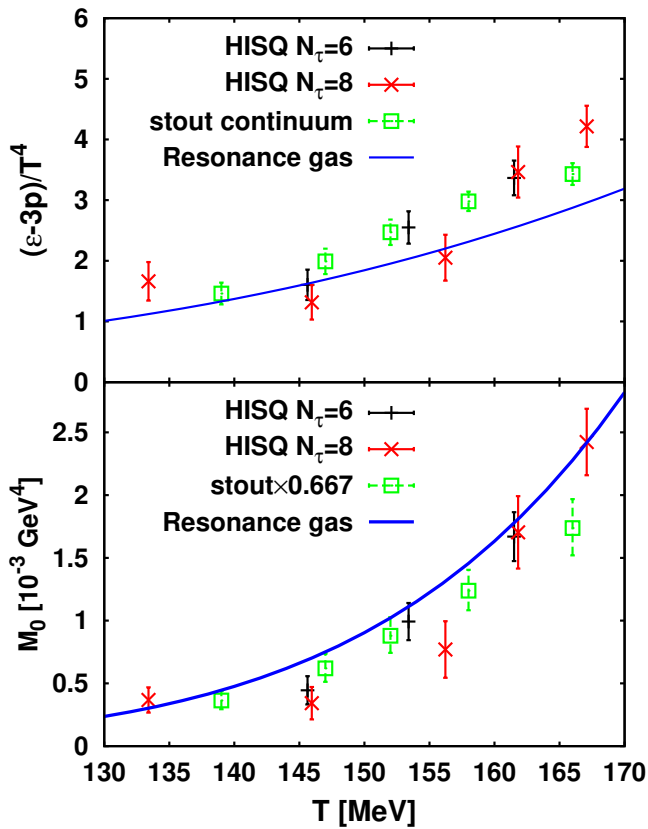}
 \caption{(Color online) Comparison of the resonance gas with lattice results. Upper: interaction measure $(\varepsilon-3p)/T^4$. Lower : its
 gluonic part $M_0$. Lattice data are taken from the HotQCD collaboration
 for HISQ action with $N_\tau=6$ and 8 \cite{bazavov10:_taste_qcd} and from the Budapest-Wuppertal
 collaboration for stout action with continuum estimation (average of
 $N_\tau=8$ and 10) \cite{borsanyi10:_qcd}. For $M_0$, stout data is
 estimated from $\varepsilon-3p$ by assuming the same ratio of the
 gluonic part as that of HotQCD. See text for details.}
 \label{fig:eos_lattice}
\end{figure}

\begin{figure}[ht]
 \includegraphics[width=3.375in]{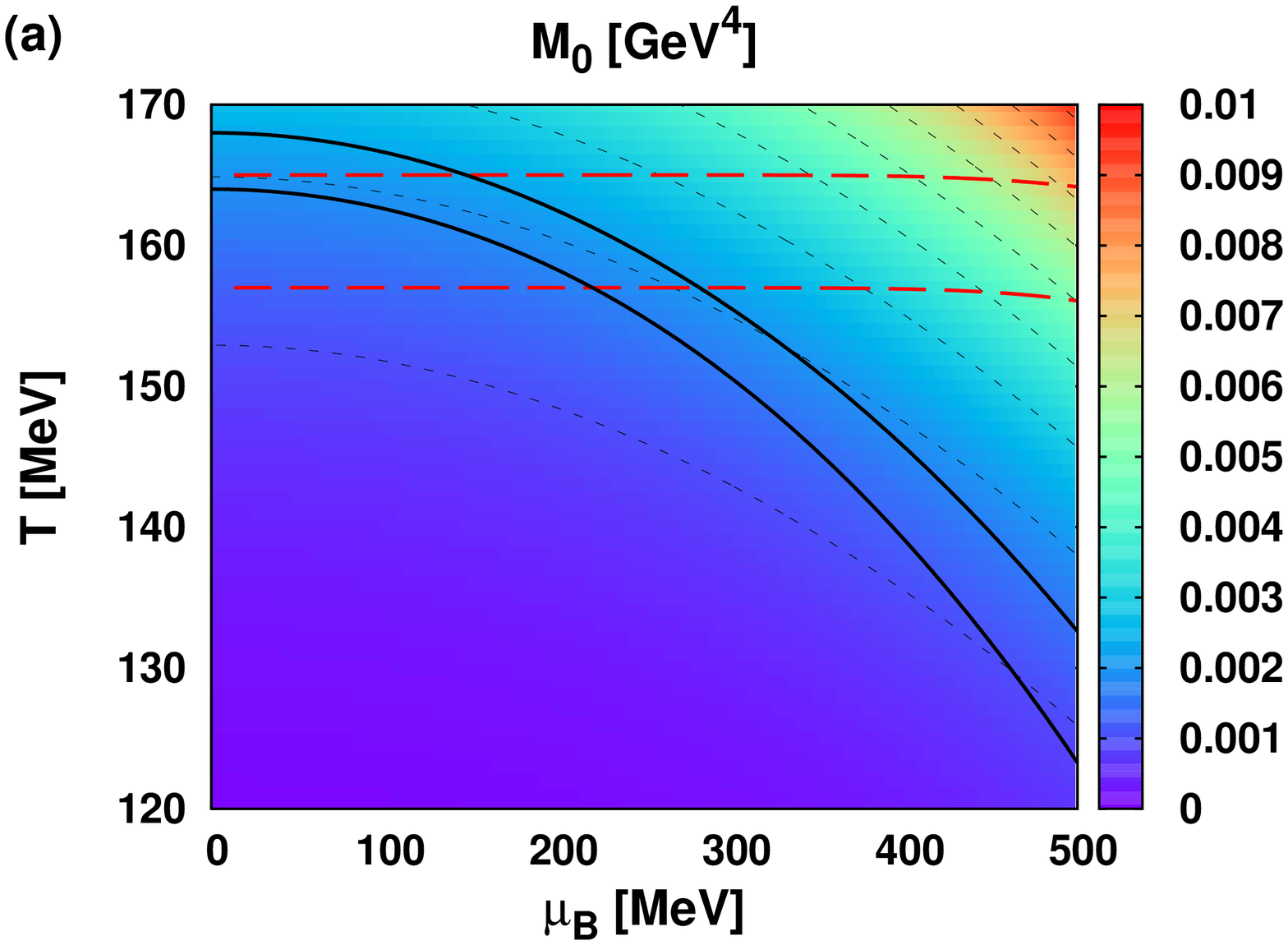}\\
 \includegraphics[width=3.375in]{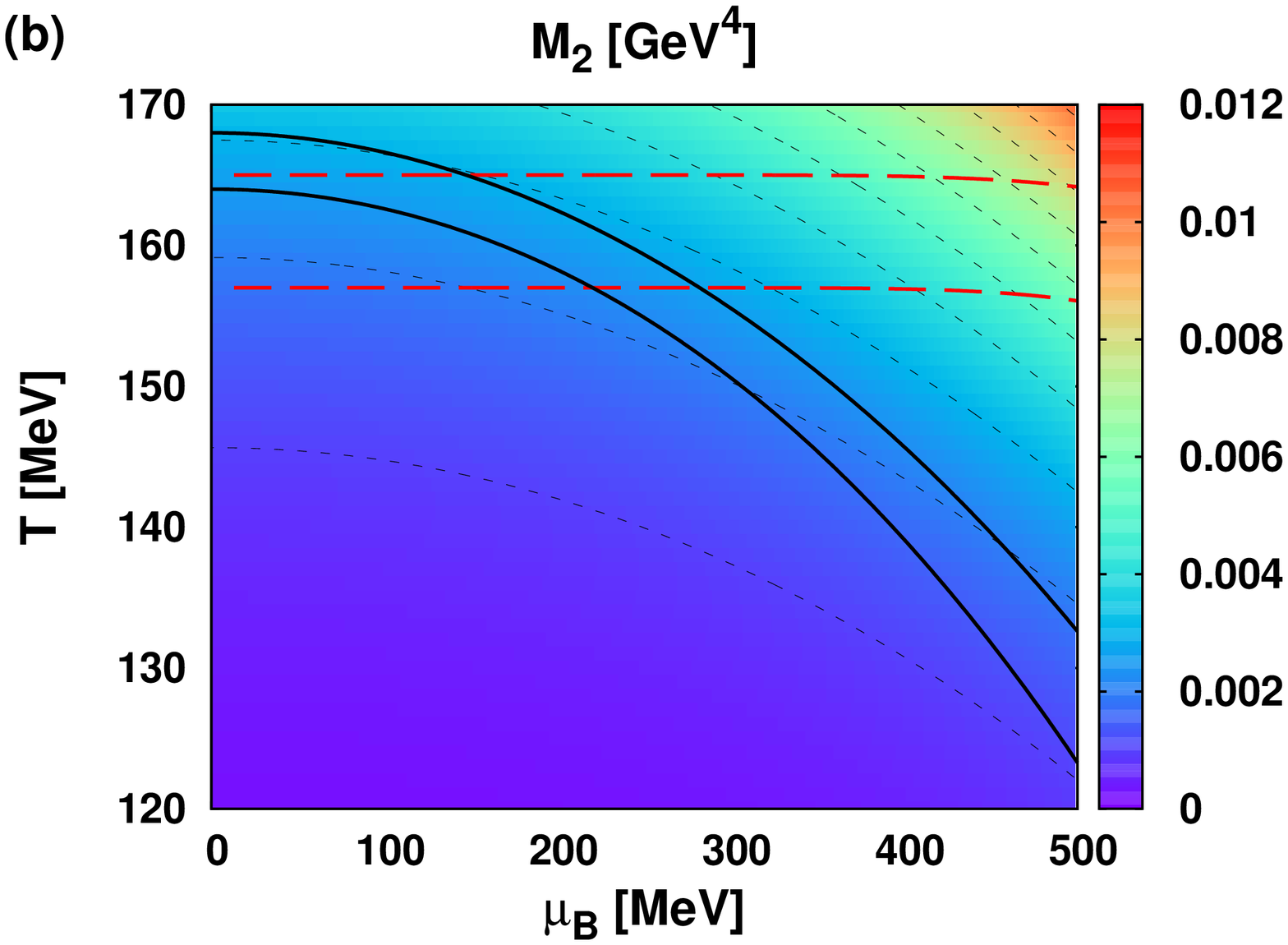}
 \caption{(Color online) Contour plots for $M_0$[(a)] and $M_2$[(b)] obtained from the resonance gas model,
 Eqs.~\eqref{eq:m0_model} and \eqref{eq:m2_model}.
 The thick solid lines (black) indicate the chemical freeze-out line including
 uncertainty of parameters given in Ref.~\cite{cleymans06:_compar}.
 The thick dashed lines (red) also indicate the freeze-out line but from \cite{andronic06:_hadron}.}
 \label{fig:m0m2-tmu}
\end{figure}

The result of $M_0$ from the resonance gas model calculated with
Eqs.~\eqref{eq:m0_model} is shown in the lower panel of
Fig.~\ref{fig:eos_lattice} together with $\varepsilon-3p$ in the upper
panel. We compare the resonance gas model with lattice data from
two different fermion discretization schemes. One is from highly
improved staggered fermion (HISQ) action, calculated by the HotQCD
collaboration with temporal extent $N_\tau=6$ and 8
\cite{bazavov10:_taste_qcd}. 
The other is from stout-link improved staggered fermion (stout) action
calculated by Budapest-Wuppertal collaboration
\cite{borsanyi10:_qcd}. In the former the light quark mass is slightly
heavier than physical one and the continuum
extrapolation is not made while the latter corresponds to
physical quark masses and gives a continuum estimation.
The upper panel shows the full trace anomaly including both gluonic and
fermionic parts for a reference.
Lattice data obtained from the different schemes show reasonable
agreement in temperature range considered here. Therefore we assume the
present HISQ data already approximates the continuum result well.
As already discussed in Ref.~\cite{borsanyi10:_qcd}, the resonance gas
model shows a small discrepancy at $T > 150$ MeV which could be
attributed to missing heavier states
\cite{majumder10:_hadron_qcd}. Since in our model $M_0$ is
essentially dominated by light hadrons, we presume that this discrepancy
does not affect the following analyses. In the lower panel, the HISQ
data shows the gluonic part of the trace anomaly.
Since the equation of state of the stout
action was calculated in a different way such that the gluonic part is
not separated \cite{borsanyi10:_qcd}, we estimate it in the following way.
 First we assume the ratio of the gluonic part
of the trace anomaly to the total one is same as those in the HISQ
data. Next, we calculate the ratio by averaging that of HISQ data for
130 MeV $< T < 170$ MeV. The upper bound corresponds to upper limit
of the pseudocritical temperature, below which we do not see clear temperature
dependency in the ratio. Then we multiply
$\varepsilon-3p$ in the stout action by the resultant ratio factor
0.667. Errors are estimated from the upper and lower value of the data
points and from the average deviation of the ratio factor 0.052 of the
HISQ data. One sees our model $M_0$ reproduces the lattice data well, and
therefore we expect that the model gives a good approximation to
$M_2$.\footnote{One may try to improve the agreement by introducing
interactions in the resonance gas. For example, we can use an excluded
volume correction to the resonance gas, which incorpolates the repulsive
interaction among hadrons \cite{rischke91}. We found, however, that $\chi^2$ fitting of
the excluded volume parameter $v_0$ to the lattice data of $M_0$ and
$\varepsilon-3p$ gives a consistent result with $v_0=0$. }
We display $M_0$ and $M_2$ as functions $T$ and $\mu_B$ in
Fig.~\ref{fig:m0m2-tmu}.

In Fig.~\ref{fig:m0m2-tmu}, we also draw the chemical freeze-out lines
proposed by two groups. One (denoted by ``FOI'') is from a combined fit to statistical
model results and has been shown to agree with various freeze-out conditions
\cite{cleymans06:_compar}. The temperature is
given by
\begin{equation}
 T(\mu_B)= a-b \mu_B^2 -c \mu_B^4 \label{eq:temperature_param1}
\end{equation}
where $a=0.166\pm 0.002$ GeV, $b=0.139\pm 0.016$ GeV$^{-1}$,  and
$c=0.053 \pm 0.021$ GeV$^{-3}$. Collision energy dependence is also
given through the chemical potential
\begin{equation}
 \mu_B(\sqrt{s})= \frac{d}{1+e\sqrt{s_{NN}}}\label{eq:chemi_param1},
\end{equation}
with $d=1.308\pm 0.028$ GeV and $e=0.273\pm 0.016$ GeV$^{-1}$.
It has been shown that this parameterization works well for recent STAR
data \cite{aggarwal11:_scalin_proper_at_freez_out}.

The other (denoted by FOII) is a parametrization of results of a statistical model
shown in Ref.~\cite{andronic06:_hadron}. The freeze-out temperature and
chemical potential are given as functions of $\sqrt{s_{NN}}$ (in unit of GeV)
\begin{align}
 T(\sqrt{s})&= T_{\text{lim}}\left( 1-
 \frac{1}{0.7+(e^{\sqrt{s_{NN}}}-2.9)/1.5} \right)
 \label{eq:temperature_param2} \\
 \mu_B(\sqrt{s}) &= \frac{a'}{1+b'\sqrt{s_{NN}}}.\label{eq:chemi_param2}
\end{align}
where $T_{\text{lim}}=161\pm 4$ MeV, $a'=1303\pm 120$ MeV and
$b'=0.286\pm 0.049$ GeV$^{-1}$.
We draw two lines for each freeze-out curves corresponding to upper and lower temperatures estimated by
uncertainties in the parameters.
The main difference between the two freeze-out curves appears $\mu_B > 300$
MeV, which corresponds to $\sqrt{ s_{NN}} < 10$ GeV. While the one by
Cleymans et al., FOI,  has a strong curvature which leads to lower freeze-out
temperature in this region, the other by Andronic et al., FOII,  shows almost
constant freeze-out temperature up to $\mu_B\sim 500$ MeV, resulting in
coincidence with the QCD chiral transition line
\cite{kaczmarek11:_phase_qcd}\footnote{Although this line corresponds to
the chiral transition, it can presumably represent the deconfinement transition line
also.}. Although the diffference between them seems to partly come from
the fact that FOI uses $4\pi$ particle yield while FOII uses
midrapidity data only, one may consider the constant temperature case
(FOII) to be simultaneous chemical freeze-out at the hadronization.

For charmonium production, it is not clear which scenario is more
likely, due to small production rate at lower energies. At the top SPS energy where
the charmonium particle ratio data is available, $\sqrt{s_{NN}}=17.3$
GeV,  the two freeze-out curves coincide.
From Fig.~\ref{fig:m0m2-tmu}, one sees both $M_0$ and $M_2$ increases
as $\mu_B$ does so at fixed temperature. This implies larger medium modification of
charmonia for larger chemical potential. If the freeze-out temperature
decreases steeper, as in FOI, however,  resultant $M_0$ and $M_2$ do not differ so
much.  Therefore, we expect that the mass shift of charmonium substantially differs
at lower collision energies between the two possible freeze-out scenarios.

\section{Mass shift of charmonium}
\label{sec:massshift}

\subsection{Second order Stark effect}

First, we calculate the mass shift of $J/\psi$ using the second order
Stark effect in QCD as done in Ref.~\cite{lee_morita_stark}. Provided
the wave function of the quarkonium in the momentum space $\psi(k)$ is normalized as
$\int\frac{d^3\boldsymbol{k}}{(2\pi)^3}|\psi(\boldsymbol{k})|^2=1$, the formula
of the mass shift for the $1S$ state is given by
\cite{peskin79,luke92,lee03,lee_morita_stark}
\begin{align}
 \Delta m_{J/\psi} &=
  -\frac{1}{18}\intop_{0}^{\infty}\frac{kdk^2}{k^2/m_c+\epsilon}\left|
							       \frac{\partial\psi(k)}{\partial k}\right|^2
 \left\langle \frac{\alpha_s}{\pi}\Delta\boldsymbol{E}^2\right\rangle_{T,\mu_B}\label{eq:stark}
 \\
 &=-\frac{7\pi^2}{18}\frac{a^2}{\epsilon}\left\langle \frac{\alpha_s}{\pi}\Delta\boldsymbol{E}^2\right\rangle_{T,\mu_B}.
\end{align}
where $k=|\boldsymbol{k}|$, $m_c$ and $\epsilon$ are the charm quark
mass and the binding energy, respectively. The above formula can be also
derived from potential non-relativistic QCD (pNRQCD), as shown in the Appendix.
The second line is obtained
for the Coulombic bound state with Bohr radius $a$. These parameters can
be determined by a fit to the $J/\psi$ mass in vacuum and the size of
the wave function in the Cornell potential model
\cite{eichten78:_charm}. It gives $m_c=1704$ MeV, $a=0.271$ fm, and
$\alpha_s=0.57$. In this formula, the mass shift is proportional to the
change of the electric condensate
$\langle \frac{\alpha_s}{\pi}\Delta\boldsymbol{E}^2\rangle_{T,\mu_B}$
from its vacuum value. The electric condensate as well as
the magnetic counterpart can be written in terms of $M_0$ and $M_2$ as
\cite{lee_morita_stark}, for $N_f=3$,
\begin{align}
 \left\langle \frac{\alpha_s}{\pi} \Delta \boldsymbol{E}^2 \right\rangle_{T,\mu_B}
  &=
 \frac{2}{9}M_0(T,\mu_B)+\frac{3}{4}\frac{\alpha_s^{\text{eff}}}{\pi}M_2(T,\mu_B),\label{eq:ec}\\
 \left\langle \frac{\alpha_s}{\pi} \Delta \boldsymbol{B}^2 \right\rangle_{T,\mu_B}
  &=
 -\frac{2}{9}M_0(T,\mu_B)+\frac{3}{4}\frac{\alpha_s^{\text{eff}}}{\pi}M_2(T,\mu_B).
\end{align}
Here the effective coupling constant $\alpha_s^{\text{eff}}$ can be
chosen according to the relevant energy scale to the expectation value
of the operator. In this case, the formula is based on OPE with
separation scale $\epsilon$. Thus it is plausible to take
$\alpha_s^{\text{eff}}=0.57$ obtained from the fit to the bound state.

\begin{figure}[!t]
 \includegraphics[width=3.375in]{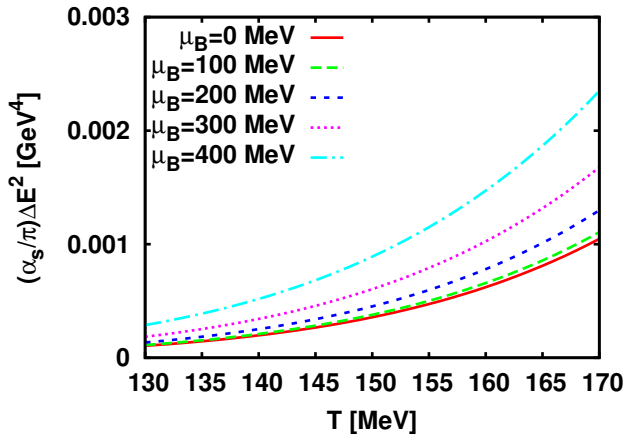}
 \caption{(Color online) Temperature dependent part of the electric condensate
 $\langle\frac{\alpha_s}{\pi}\Delta \boldsymbol{E}^2\rangle$. Each line
 stands for the case of different chemical potential.}
 \label{fig:ec}
\end{figure}

\begin{figure}[!t]
 \includegraphics[width=3.375in]{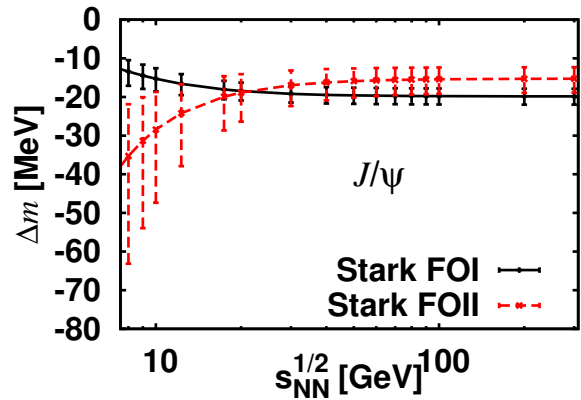}
 \caption{(Color online) Mass shift of $J/\psi$ at freeze-out temperature and chemical
 potential from the second order Stark effect. The horizontal axis
 denotes the collision energy which is
 related to the temperature and the chemical potential via
 Eqs.~\eqref{eq:temperature_param1} and \eqref{eq:chemi_param1} for the FOI
 case and Eqs.~\eqref{eq:temperature_param2} and \eqref{eq:chemi_param2}
 for the FOII case.}
 \label{fig:stark}
\end{figure}

Figure \ref{fig:ec} shows the electric condensate for
$\mu_B=0, 100,200, 300$ and 400 MeV as a function of
temperature. The maximum chemical potential 400 MeV roughly corresponds
to 40$A$ GeV Pb+Pb collisions at SPS, the lowest collision energy above
$J/\psi$ production threshold \cite{andronic08}.
One sees larger change of the electric condensate for high temperature
and chemical potential as is expected from Fig.~\ref{fig:m0m2-tmu}.
 Now we are able to estimate the mass shift of $J/\psi$ by
putting the condensate into Eq.~\eqref{eq:stark}. We compute the mass
shift along the freeze-out lines shown in Fig.~\ref{fig:m0m2-tmu} and
show the result as a function of $\sqrt{s_{NN}}$ in Fig.~\ref{fig:stark}
For higher colliding energies than $\sqrt{s_{NN}} > 30$ GeV, the mass
shift is independent of the colliding energy owing to the fact that the
freeze-out temperature varies little and the chemical potential does not
change the condensate significantly (see Fig.~\ref{fig:ec}.)
The amount of the downward mass shift is 10--20 MeV, including uncertainty.
On the other hand, low energy results differ between FOI and FOII, as
expected from Fig.~\ref{fig:m0m2-tmu}. Along the FOI chemical freeze-out
line, the mass shift becomes smaller while opposite behavior is seen
for FOII. In the FOII case, since temperature is almost constant, the
larger the chemical potential, the bigger the mass shift becomes owing to the effects from the
chemical potential. In the other case, however, decreasing freeze-out
temperature cancels the effect of the chemical potential. For instance,
one can see in Fig.~\ref{fig:ec} that $\Delta \boldsymbol{E}^2$ at
$T=165$ MeV and $\mu_B=0$ is almost equal to that at $T=145$ MeV and $\mu_B=400$
MeV. Therefore,
At the lowest SPS energy, the downward mass shift ranges  from 10--60 MeV,
depending on the choice of the thermal parameters.

\subsection{QCD sum rules}
\label{sec:qcdsr}

While the second order Stark effect provides the downward mass shift
directly in the case of the increasing electric condensate, QCD sum rule
is expected to be more quantitatively reliable according to the larger
separation scale by going to the deep Euclidean region.
Here we give an estimation based on the Borel sum rule framework
used in Ref.~\cite{morita_borel} with further improvement as described below.

The two point current correlation function in the vector channel after
the Borel transformation with the Borel mass $M^2$ is given by
\cite{bertlmann82,Furnstahl_PRD42,morita_borel}
\begin{align}
 \mathcal{M}(M^2) &= e^{-\nu}\pi
  A(\nu)[1+\alpha_s(M^2)a(\nu)+b(\nu)\phi_b(T) \nonumber\\
 &+c(\nu)\phi_c(T)].
 \label{eq:borel_moment}
\end{align}
with $\nu=4m_c^2/M^2$. Here the system is assumed to be at rest with
respect to the medium.
The Wilson coefficients $a(\nu)$, $b(\nu)$, and
$c(\nu)$ are listed in Ref.~\cite{morita_borel}. The temperature
dependency is governed by the dimension four gluon condensate terms
$\phi_b(T)$ and $\phi_c(T)$ given by
\begin{align}
 \phi_b&= \frac{4\pi^2}{9(4m_c^2)^2}G_0(T),\label{eq:phib}\\
 \phi_c &= \frac{4\pi^2}{3(4m_c^2)^2}G_2(T),\label{eq:phic}
\end{align}
The relations of the positive definite quantities $M_0$ and
$M_2$ to the dimension four gluon condensates appearing in the OPE side
correlation function are given by
\begin{align*}
 G_0(T)&=  G_0^{\text{vac}}-\frac{8}{9}M_0(T)\\
 G_2(T)&= -\frac{\alpha_s^{\text{eff}}}{\pi} M_2(T)
\end{align*}
after taking the one-loop expression for the beta function.
As in the pure gauge case, we use a temperature dependent effective
coupling constant $\alpha_s^{\text{eff}}=\alpha_{qq}(T)$
extracted from lattice calculation of the color singlet heavy quark free energy by
assigning $\alpha_s\langle G^a_{\mu\rho}G^{a\nu\rho}\rangle_T \equiv
\langle \alpha_s(T)G^a_{\mu\rho}G^{a\nu\rho} \rangle$ in the spirit of
the separation scale in the heavy quark system which imposes all the
tempeture effect on the condensates \cite{lee_morita_stark,Hatsuda93}.
We take the values from $N_f=2$ results of $\alpha_{qq}(r_{\text{max}})$
in Fig.~6 of Ref.~\cite{kaczmarek05:_static_qcd}.
To account for different critical temperatures between the $N_f=2$
simulation, $T_c=202$ MeV,  and the reality (see Sec.~\ref{sec:intro}),
we rescale the coupling constant read off from the data by assuming similar temperature
dependency to that of $T_c=170$ MeV. This value is considered to be the
upper bound of the pseudocritical temperature in reality, owing to
measurement based on the strange quark number susceptibility
$T_{pc}=165(5)(3)$ MeV in Ref.~\cite{borsanyi10:_in_t_qcd}

Presumably, this choice does not affect the results
quantitatively since the twist-2 contribution to the medium modification
is relatively small in the hadronic phase \cite{morita_jpsifull}.
The $T=0$ part of the scalar gluon condensate
$G_0^{\text{vac}}$ is fixed to be $(0.35 \text{GeV})^4$
\cite{reinders85} as in our previous
calculations.\footnote{$G_0^{\text{vac}}$
has still a large error after fitting to the various experimental data, see also
\cite{ioffe06:_qcd_at_low_energ} for example. This vacuum value, however, does not
affect the in-medium effect significantly  since we are looking at
relative changes from vacuum.}


The Borel-transformed correlation function is related to the spectral
density through the dispersion relation
\begin{equation}
 \mathcal{M}(M^2) =
  \int_{0}^{\infty}ds\,e^{-s/M^2}\text{Im}\tilde{\Pi}(s). \label{eq:dispersion}
\end{equation}
We model the right-hand side of the dispersion relation with a
simple ansatz and call it the phenomenological side as usual. In the
hadronic medium
below $T_c$, previous analyses based on the gluon condensate of pure
gauge theory indicates the broadening is small enough to ignore.
Provided the continuum part of the model spectral density
$\mathcal{M}^{\text{cont}}(M^2)$, the the mass of $J/\psi$ is given by
 \begin{gather}
 m_{J/\psi}^2(M^2)=
  -\frac{
   \frac{\partial}{\partial(1/M^2)}
   [\mathcal{M}(M^2)-\mathcal{M}^{\text{cont}}(M^2)]}
   {\mathcal{M}(M^2)-\mathcal{M}^{\text{cont}}(M^2)}.\label{eq:ordinary_sumrule}
 \end{gather}
We use the perturbative expression up to $\mathcal{O}(\alpha_s)$ for
$\mathcal{M}^{\text{cont}}(M^2)$ as in Ref.~\cite{morita_borel}.

\begin{figure}[ht]
 \includegraphics[width=3.375in]{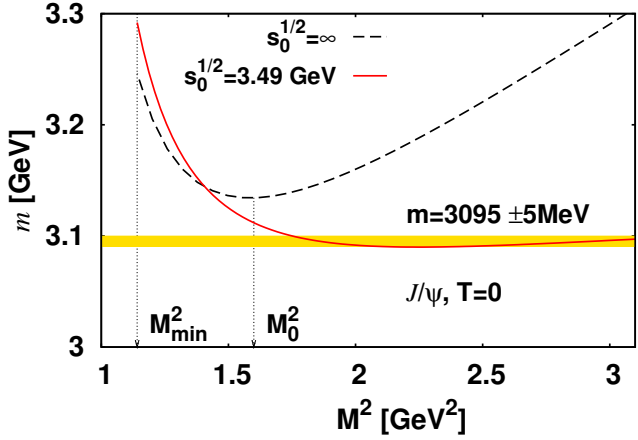}
 \caption{(Color online) Borel curves at $T=0$. The dashed line stands for case of
 $\sqrt{s_0}=\infty$. The solid line denotes the case of
 $\sqrt{s_0}=3.49$ GeV which gives the flattest curve according to
 Eq.~\eqref{eq:deltam}. The band indicates the systematic uncertainty
 associated with the flattest Borel curve.}
 \label{fig:borelT0}
\end{figure}

Since the mass of $J/\psi$ is a function of the Borel mass $M^2$ which
is an unphysical parameter, one has to choose the range of $M^2$ called
Borel window by following the criteria;
\begin{enumerate}
 \item $M^2_{\text{min}}$ : Convergence of the OPE by imposing the dimension four operator
       contribution less than 30\% to the total OPE
       \cite{Shifman_NPB147}.

 \item $M^2_{\text{max}}$ : Continuum contribution to the dispersion
       integral is less than 30\%. We choose the threshold parameter
       $s_0$ such that extracted $J/\psi$  mass is least sensitive to $M^2$.
\end{enumerate}
As discussed in Ref.~\cite{morita_borel}, the values 30\% are physically
reasonable but arbitrary. Due to truncation of the OPE, we cannot obtain the
completely $M^2$ independent mass. Specifically the mass strongly varies
with $M^2$ at lower $M^2$ even inside the Borel window. As this can be
regarded as a systematic uncertainty due to the truncation, we take this
effect into account in the mass evaluation by averaging the mass over
the Borel window and take its variance as the error \cite{Hatsuda93}.
Namely,
\begin{equation}
 \bar{m} = \left.\int_{M^{2\prime}_{\text{min}}}^{M^2_{\text{max}}} dM^2
  m(M^2)\right/ (M^2_{\text{max}}-M^{2\prime}_{\text{min}})\label{eq:avm}
\end{equation}
and
\begin{equation}
 (\delta m)^2 = \left.\int_{M^{2\prime}_{\text{min}}}^{M^2_{\text{max}}} dM^2
  (m(M^2)-\bar{m})^2 \right/ (M^2_{\text{max}}-M^{2\prime}_{\text{min}}).\label{eq:deltam}
\end{equation}
An example taken from $T=0$ is shown in Fig.~\ref{fig:borelT0}.
Here the Borel window is defined by $M^2 \in
[M^{2\prime}_{\text{min}},M^2_{\text{max}}]$ and
$M^{2\prime}_{\text{min}}\equiv \text{max}(M^2_{\text{min}},M_0^2)$.

We have introduced $M_0^2$ such that
$dm(M^2;\sqrt{s_0}=\infty)/dM^2 =0$ in order to remove the strongly
$M^2$ dependent part of $m(M^2)$ from the evaluation of the average
\eqref{eq:avm} and variance \eqref{eq:deltam}. This is a reasonable
choice as the continuum threshold is so determined that it makes the
Borel curve flattest at $M^2 > M_0^2$.

Figure \ref{fig:borelT0} shows an example of the determination process.
We start with $\sqrt{s_0}=\infty$ case which gives $M_0^2$ shown as the dashed line.
Then we search for $\sqrt{s_0}$ such that it gives the smallest $\delta m$
using Eq.~\eqref{eq:deltam} which takes its minimum when the deviation
from the average value is the smallest. The resultant average deviation
is indicated by the band in the figure. Irrespective to the temperature,
it is found to be approximately 5 MeV and 6 MeV in the case of $J/\psi$
and $\chi_{c1}$, respectively.
Parameters of the theory are fixed to $m_c(p^2=-2m_c^2)=1.262$ GeV and
$\alpha_s(8m_c^2)=0.21$ by fitting to the vacuum $J/\psi$ and
$\chi_{c1}$ masses on the basis of the same criterion of the Borel
window. This process removes the ambiguity on the arbitrary choice of
the criterion mentioned above.
Including the width is straightforward. As shown
in Ref.~\cite{morita_borel}, introducing width increases the mass at
small $M^2$. This fact leads to larger $s_0$ after minimizing
$\delta m$. Then we will have the mass-width relation similar to those
shown in Refs.~\cite{morita_jpsiprl,morita_jpsifull,morita_borel}.
In most cases
$M_0^2 > M^2_{\text{min}}$ holds in the charmonium sum rules.
At high temperature and chemical potential, however, we found that
the Borel stability is lost \cite{morita_jpsiprl,morita_borel} thus
$M_0^2$ is not well defined. In such cases, we can still recover the Borel
stability by decreasing the threshold parameter or by introducing the
width. When the width must be introduced, we cannot determine both mass
and width simultaneously but have only constraints. In what follows, we
restrict ourselves to cases in which the Borel stability is established
with vanishing width.
\begin{figure}[ht]
 \includegraphics[width=3.375in]{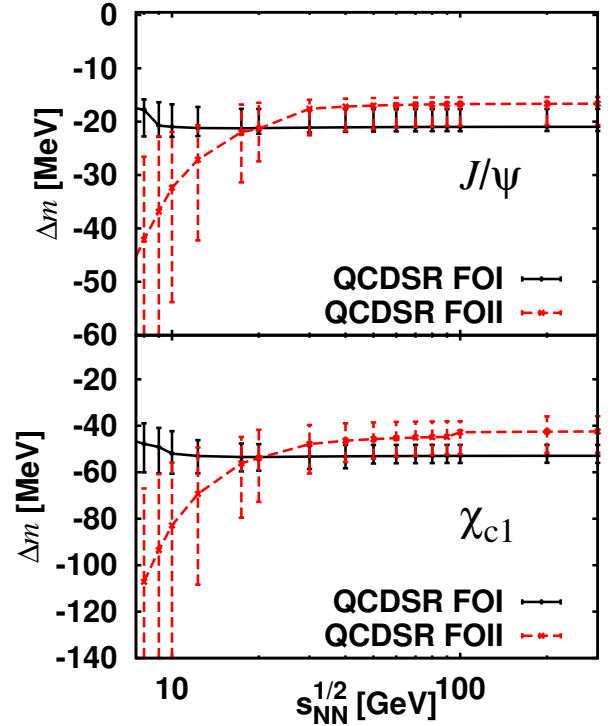}
 \caption{(Color online) Mass shift of $J/\psi$ (upper) and $\chi_{c1}$
 (lower) obtained with QCD sum rules. Solid and dashed lines stand for
 the mass shift corresponding to the different freeze-out curves, as in
 Fig.~\ref{fig:stark}. Errors are calculated from the uncertainty in the
 thermal parameters. }
 \label{fig:massshift_sumrule}
\end{figure}

Figure \ref{fig:massshift_sumrule} displays the results of the mass
shift obtained from the QCD sum rule analysis as described above.
We plot the mass shifts corresponding to the two freeze-out curves as in
the Stark effect results (Fig.~\ref{fig:stark}). Errors are estimated
from the uncertainty in the thermal parameters in
Eqs.~\eqref{eq:temperature_param1}--\eqref{eq:chemi_param2}.
The systematic errors in the Borel sum rules which have been introduced
above are not included in the plot. We also calculate mass shift of
$\chi_{c1}$ in the same way. In $\chi_c$, we have observed the loss of
the Borel stability at lower collision energies than $\sqrt{s_{NN}} < 8$
GeV in FOII, owing to much change of the gluon condensate
\cite{morita_borel}.
 Comparing the result with Fig.~\ref{fig:stark}, one finds that the two
 methods give consistent mass shifts as having been found in
 Ref.~\cite{morita_borel}.
The mass shift of $\chi_{c1}$ is approximately
twice as large as that of $J/\psi$, as previously found
\cite{song09,morita_borel}.
We expect similar results for other $\chi_c$ states.

Before closing the section, we would like to comment on the effect of
the scattering term, which was pointed out in Ref.~\cite{Bochekarev86}
to appear as a pole at the zero mode in the correlator.
In our previous works, it was neglected since such a contribution
appears in the OPE side to cancel the phenomenological side put as an
ansatz. In the deconfined phase, this argument should hold because the
physical particle absorbing the current is the (anti)charm
quark, while this is not so in the hadronic phase where charmed
mesons are the physical particles. In general, these terms can be
neglected as they contribute at zero energy in the spectral density.
However, without invoking such arguments, it can be neglected in the
present case on the following grounds.
First, the scattering terms in the OPE and the phenomenological side
will be proportional to $e^{-m_c/T}$ and  $e^{-m_D/T}$ respectively,
while the other OPE terms in the Borel transformed sum rule will in
general scale as $e^{-4m_c^2/M^2}$.  Hence as long as $T <M^2/(4m_c)$ or
$T<M^2 m_D/(4m_c^2)$, the scattering terms can be neglected.  Since the
the smallest Borel mass relevant in our analysis is always larger than 1
GeV, taking $m_c=1.26 $ GeV, one finds that the scattering terms can be
safely neglected for $T < 200$ MeV.  Moreover, the open
charm meson will receive greater medium effect than charmonia, making
$m_D$ close to $m_c$. Thus the scattering contribution from the OPE and
the phenomenological side will tend to cancel each other.
Second, one can remove the scattering contribution by making use
of the fact that it contributes as a constant term to the Borel transformed
correlator $\mathcal{M}(M^2)=\int ds e^{-s/M^2} \rho(s)$ since the scattering
contribution to the spectral density takes a form as $\rho^{\text{scat}}(s)\propto\delta(s)$
\cite{Bochekarev86}(equivalently $\sqrt{s} \delta(\sqrt{s})$
\cite{aarts05,mocsy06,umeda07}). Therefore, the effect of the scattering
term does not exist in the derivative of $\mathcal{M}(M^2)$ with respect
to $M^2$ ($1/M^2$ in practical calculations).  One may then start from
the once differentiated sum rule and express the mass in the $\Gamma=0$ limit as
\begin{equation}
 m^2_{c\bar{c}} = \frac{ \frac{\partial^2}{\partial (1/M^2)^2}
  [\mathcal{M}(M^2)-\mathcal{M}^{\text{cont}}(M^2)]}{-\frac{\partial}{\partial(1/M^2)}[\mathcal{M}(M^2)-\mathcal{M}^{\text{cont}}(M^2)]}.\label{eq:deriv_sumrule}
\end{equation}
Indeed such a method was advocated in Ref.~\cite{koike95:_qcd} but
criticized in Ref.~\cite{hatsuda95:_qcd} because starting with a higher
order derivative make the OPE side more sensitive to unknown higher
dimensional condensates and Borel stability is lost;
it was claimed in Ref.~\cite{hatsuda95:_qcd} that due to this artifact
the light vector meson was found to increase in the medium in
Ref.~\cite{koike95:_qcd}.
This comes from the fact that the OPE in the light vector meson has no
scale parameter other than the Borel mass; therefore after the Borel
transformation the sum rule becomes a polynomial in $1/M^2$ with the
highest power determined by the highest dimensional operator calculated
in the  OPE. In the case of heavy quarkonia, however, the presence of
heavy quark mass does not make the OPE a mere polynomial in $1/M^2$ so
that stability is not lost even after derivatives.

\begin{figure}[t]
 \includegraphics[width=3.375in]{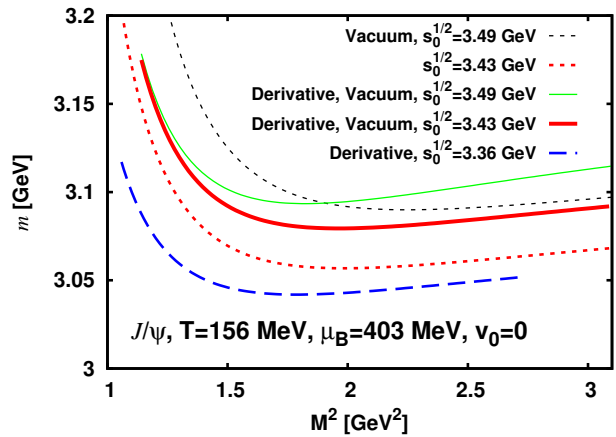}
 \caption{(Color online) Borel curves for the mass obtained from the ordinary QCD sum rule
  and from the differentiated one, Eq.~\eqref{eq:deriv_sumrule}. See
 text for detailed explanation.}
 \label{fig:borel}
\end{figure}

Figure \ref{fig:borel} shows an evidence for the above argument. We show
the Borel curves obtained from
Eq.~\eqref{eq:deriv_sumrule} as well as those from the ordinary method,
Eq.~\eqref{eq:ordinary_sumrule}.
For illustration, we show a case with large medium effect, corresponding
to $\sqrt{s_{NN}}=8.7$ GeV in FOII, of which temperature and chemical
potential are $T=156$ MeV and $\mu_B=403$ MeV, respectively.
For vacuum, we display three curves. The thin dotted (black) shows the
same one as in Fig.~\ref{fig:borelT0} for the reference. The thin solid
(green) denotes that obtained from Eq.~\eqref{eq:deriv_sumrule} with the
same value of $\sqrt{s_0}$.
One sees both curves give almost the same mass. After minimizing
$\delta m$ in the differentiated sum rule
[Eq.~\eqref{eq:deriv_sumrule}], one gets a slightly smaller mass
indicated by the thick solid (red) line. One notes
Eq.~\eqref{eq:deriv_sumrule} gives smaller
mass at small $M^2$, as expected from the fact that it becomes sensitive
to higher dimensional operators. The small discrepancy of the mass
can be attributed to the dimension six contribution which slightly
increases the mass~\cite{Kim01}.
The remaining two curves are for the
medium. The thick dotted (red) is obtained from the ordinary sum rule
while the thick dashed (blue) is from the derivative with the
optimization.  If one compares them with the corresponding results for
vacuum, one finds that the mass shifts are almost the same between the two
sum rules. Hence, we conclude the general properties of the in-medium
modification of heavy quarkonia at low temperature up to near
$T\simeq 160$ MeV will
not change by including the scattering contribution.

\section{Implication for experiments}
\label{sec:imp_exp}

We have discussed mass shifts of $J/\psi$ in hot medium, which could be
produced in heavy ion collisions. Production mechanism of $J/\psi$ has
not been fully understood yet, because of the still unknown elementary production process
and also the complicated collision processes
\cite{rapp:_charm}. In the following we briefly review the collision
process and specify the situation we will consider in this section.
Charmonia have been considered to be mostly produced by collisions
between initial state quarks and gluons in nuclei at the initial stage of
heavy ion collisions. The formation time scale $\tau \sim 1/(2m_c)$ is
supposed to be shorter than the thermalization time scale of the medium, which is related
to flow measurements through hydrodynamic model calculations.
The produced charmonia will also interact with colliding nuclei. The
dissociation of $J/\psi$ by this interaction, called cold nuclear
matter effect, is estimated by the nuclear absorption cross section,
which is roughly 1.5 mb at RHIC energies and 4.4 mb at SPS energies
\cite{zhao08:_trans_momen_spect_of_j}. In the hot medium, charmonia
could melt.
While Lattice QCD have shown existence of the spectral peak even at
higher temperature \cite{Asakawa_PRL92,Datta_PRD69,jakovac07,aarts07},
model calculations can explain the lattice data with melting of $J/\psi$
\cite{mocsy08}. Even if the bound states can survive, they will acquire
substantial collisional broadening through interacting with quarks and gluons in medium
\cite{rapp:_charm,Park}. There could be also recombination of a $c\bar{c}$
pair inside deconfined medium below dissociation temperature. Finally,
(anti-)charm quarks hadronize at the phase boundary to form charmed
mesons, baryons, and hidden charm states. If $J/\psi$ mass is modified
in the medium and decay \textit{inside} the mediun, one may be able to
observe it as modification of the peak in dilepton channel.
However, one needs a dynamical approach in order to take into account
various process described above, as well as microscopic information like cross sections
 to estimate the final yield \cite{hashimoto86,rapp:_charm,Song:2010ix}.

Here we consider an alternative possibility of indirect observation
via statistical production
\cite{gazdzicki99:_eviden_j_gev,andronic03:_statis_sps_rhic_lhc,andronic07_plb652,andronic08,andronic09:_statis}.
This has been already considered as a part of the contribution in the
transport approach and could give substantial
contribution to the final yields at high centrality \cite{rapp:_charm}.
In Ref.~\cite{andronic08}, in-medium effect on the charmonium yields was
considered as a result of $D$ meson mass modification and charm
conservation. As emphasized in the literature, charm conservation plays
an important role in the statistical description of charm quarks. However, as for
the particle ratio between charmonia, the fugacity factor cancels and
the ratio can be expressed as that of thermal number densities at  given
temperature and chemical potential. Therefore,
we can focus on observables dominated by the statistical production.

First we examine $N_{\psi'}/N_{J/\psi}$ at midrapidity which was also
investigated in Ref.~\cite{andronic09:_statis}, since there was
experimental data for Pb+Pb collisions at $\sqrt{s_{NN}}=17.3$ GeV in
CERN-SPS. The temperature and chemical potential in the two freeze-out
curves coincide at this enegy as seen from Fig.~\ref{fig:m0m2-tmu}. We
adopt $T=160$ MeV and $\mu_B=240$ MeV in the following
\cite{andronic09:_statis} and discuss possible effects of charmonium
mass shifts.

Although the QCD sum
rule method cannot assess in-medium modification of $\psi'$ \footnote{In
the Borel transformed sum rule, we can obtain the same $J/\psi$ mass
even if we incorpolate the $\psi'$ contribution in the model spectral function
explicitly \cite{morita_borel}. Therefore, in-medium effect on $\psi'$
is incorpolated in the effective threshold parameter $\sqrt{s_0}$.
This shows the strong sensitivity of the sum rule to the lowest pole of
the spectral function.}, it is
expected to be strongly affected \cite{Karsch06}.
While applicability of the formula of the second order Stark effect in
Ref.~\cite{peskin79} is questionable for the physical $\psi'$ , a
crude estimation based on the dipole nature
might be possible. If we assume the mass shift scales with the size of
the wave function, the mass shift of $\psi'$ is a factor 4.2 larger than
that of $J/\psi$. For the SPS data at $\sqrt{s_{NN}}=17.3$ GeV which we
will analyze below, $\Delta m_{\psi'}$ becomes 63--119 MeV by adopting
the result from QCD sum rules and fully taking the errors into account.
Here, we vary $\psi'$ mass shift in a broader range than the above
estimation and calculate the ratio as a function of
$\Delta m_{\psi'}$. The total number of $J/\psi$ is obtained by summing
up decay contributions from $\psi'$ and $\chi_{cJ} (J=0,1,2)$.
We assume the mass shifts of different $\chi_c$ states to be the same as
that of $\chi_{c1}$ and that the branching ratios to be the same as
their vacuum values.  Specifically, the branching ratios of $\psi'$ and
$\chi_{cJ} (J=0,1,2)$ to $J/\psi$ are taken to be 0.595, 0.0116, 0.344
and 0.195, respectively \cite{pdg2010}.

\begin{figure}[t]
 \includegraphics[width=3.375in]{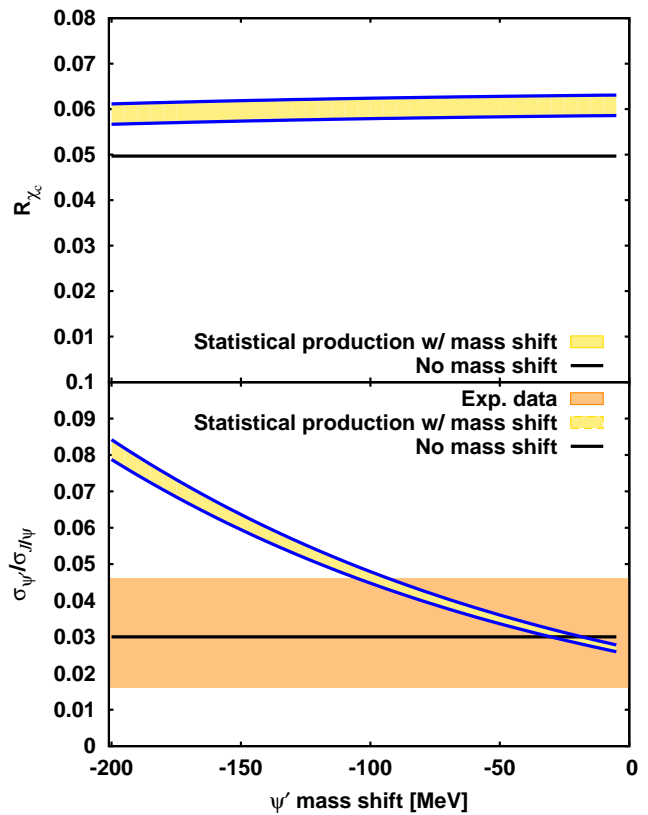}
 \caption{(Color online) Particle number ratio at $T=160$ MeV and $\mu_B=240$ MeV as a function of $\psi'$
 mass shift. The band in the lower panel indicates the experimental data
 taken from Pb+Pb collisions at $\sqrt{s_{NN}}=17.3$ GeV measured by
 NA50 collaboration \cite{alessandro05:_j_pb_pb_gev}. The solid lines
 stand for the results without any mass shift in the all charmonia for comparison.}
 \label{fig:ratio}
\end{figure}

We plot the result for $T=160$ MeV and $\mu_B=240$ MeV
together with experimental data taken from Pb+Pb collisions at
$\sqrt{s_{NN}}=17.3$ GeV \cite{alessandro05:_j_pb_pb_gev} in the lower
panel of Fig.~\ref{fig:ratio}.
One sees that a strong variation of the ratio as a function of $\psi'$
mass shift. The band for our result corresponds to the uncertainty in
the mass shift shown in Fig.~\ref{fig:massshift_sumrule}.
If the $\psi'$ downward mass shift is smaller than 100 MeV, which is a
reasonable value according to the crude estimation, the expected
mass shifts are consistent with the experimental data. At present,
however, it seems difficult to draw a conclusion from this observation;
in-medium modification of $\psi'$ is still controversial. While the
ground state, $J/\psi$, seems to exist in the vicinity of the transition,
$\psi'$ can be dissociated at finite temperature even below transition
temperature according to potential model
calculations \cite{mocsy:_poten}. Even if $\psi'$ is statistically
produced with the large mass reduction, it can still be dissolved into charmed
mesons in hadronic medium thus observed yields might be smaller than
what was produced at the hadronization: the dissociation cross section
of $J/\psi$ in hadronic matter is expected to be smaller than that of
$\psi'$ due to its smaller size~\cite{Song:2010ix}.
In this case, there must be of course enhancement of charmed mesons due
to charm conservation.
The analysis might become complicated if initially produced charmonium
could survive the quark-gluon plasma phase and the hadron phase.  Then
there will be modification to statistical model prediction as discussed
in a two component models of charmonium production in a heavy ion
collision~\cite{Grandchamp:2002wp,Song:2010ix}.  However, even within the two
component model, the thermal production will be dominant at LHC and
production ratios can reveal vital information at the hadronization
point.

Although experimentally challenging, more promising observable could be a
ratio of $\chi_c$ to $J/\psi$. In the upper panel, we display
a fraction of $J/\psi$ coming from $\chi_{c1}$ and $\chi_{c2}$ to the
total number of $J/\psi$ for
$T=160$ MeV. This quantity, $R_{\chi_c}$ was measured by HERA-B Collaboration in
proton-nucleus collisions \cite{hera-b_chic}. Our result shows
10--20\% increase of this quantity for the statistical production with
mass reductions, almost independent of the uncertain $\psi'$ mass shift.

\section{Summary}
\label{sec:summary}

In this paper, we investigate the mass shift of charmonia induced by change of
gluon condensates in hadronic medium by making use of both perturbative
(QCD second order Stark effect) and non-perturbative (QCD sum rule)
approaches. The inputs for the medium effect is the gluon condensates
calculated by a resonance gas model that
well reproduces the thermodynamic quantities calculated on the lattice
QCD result in the temperature region considered here. Extending it to
the finite baryonic chemical potential, we found that the change of the
gluon condensates becomes larger at large chemical potential, which
could result in a larger medium effect on charmonium production in heavy
ion collisions at lower colliding energies, depending on hadronization
temperature.
We found that both the perturbative and non-perturbative estimations
give almost the same mass shift. After elaborating on the sum rule
analysis, we estimated systematic error on the obtained mass shift and
found that most of the uncertainty comes mainly from thermal parameters.
While the mass shift is almost constant along the chemical freeze-out line of which
freeze-out temperature decreases as chemical potential increases (FOI),
it can exhibit stronger downward shift at lower colliding energies if
the charmonia are produced at hadronization and simultaneously frozen
out (FOII).

We consider experimental implications of this observation in the context
of statistical hadronization picture in which hadronization and the
freeze-out of charmonium occur simultaneously.
We found the data in Pb+Pb collisions at $17A$GeV are consistent with
mass reduction of $J/\psi$ and $\chi_c$ although the effect on $\psi'$
should be clarified before final conclusion is made. We pointed out
that 10--20\% enhancement of the production ratio between $\chi_c$ and
$J/\psi$  could be a signature of the downward mass shift, which
indicate a precursor of the deconfinement phenomenon.

\begin{acknowledgments}
 K.M. would like to thank K.~Redlich and W.~Weise for fruitful
 discussions during YIPQS international workshop on ``New Frontiers in
 QCD 2010''.
 This work was supported by Korean Ministry of Education through the
 BK21 Program and KRF-2011-0020333. K.M.'s work is supported by Frankfurt
 institute of Advanced Studies and the YIPQS program at Kyoto
 University.
 The work of SH was supported in part by the Asia Pacific Center for
 Theoretical Physics through the topical research program "New perspectives
 on sQGP".
\end{acknowledgments}

\appendix*

\section{Derivation of the second order Stark effect formula from Potential NRQCD}

In this appendix, we show that the second order Stark effect,
Eq.~\eqref{eq:stark}, can be derived from potential non-relativistic QCD (pNRQCD)
which provides a systematic perturbative approach to the OPE.

\begin{widetext}
Here the effective lagrangian in the static limit is given by~\cite{brambilla00:_poten_nrqcd,Brambilla05},
\begin{align}
{\cal L} & = -\frac{1}{4} G^a_{\mu \nu}G^{a \mu \nu}+ \sum_{i=1}^{n_f} \bar{q}_i iD\hspace{-0.5em}/ q_i
 + \int d^3r \text{Tr}\bigg\{ \text{S}^\dagger \bigg[
 i\partial_0 +C_F \frac{\alpha_{V_S}}{r} \bigg] \text{S} +
 \text{O}^\dagger \bigg[iD_0-\frac{1}{2N_c} \frac{\alpha_{V_O}}{r}
 \bigg] \text{O} \bigg\} \nonumber \\
 &+V_A \text{Tr} \{ \text{O}^\dagger \vec{r} \cdot
 g\vec{E} \text{S} + \text{S}^\dagger \vec{r} \cdot g \vec{E}\text{O} \}
 +
 \frac{V_B}{2} \text{Tr} \{ \text{O}^\dagger \vec{r} \cdot g
 \vec{E}\text{O}
 +\text{O}^\dagger \text{O} \vec{r} \cdot g \vec{E}  \}+ \cdots.
\end{align}
The fields $\text{S}=S \mathbf{1}_c/ \sqrt{N_c}$ and
$\text{O}=O^a T^a/\sqrt{T_F}$
are normalized static quark-antiquark singlet and octet
fields, respectively, $\vec{E}$ is the chromoelectric field, and
$C_F=(N_c^2-1)/(2N_c)=4/3$.  The trace is over the color indices.  The
matching coefficients at leading order are, $\alpha_{V_S}=\alpha_s,
\alpha_{V_O}=\alpha_s, V_A=1, V_B=1$.  The leading order correction to
the singlet potential at finite temperature as given in Eq. (64) of
Ref. \cite{Brambilla08}  is,
\begin{eqnarray}
[\delta V_S(r)]_{11} =-ig^2 \frac{T_F}{N_c} \frac{r^2}{d-1}
 \int_0^\infty dt \, e^{-it\Delta V}\,\left[\left\langle \vec{E}^a(t)
 \phi(t,0)_{ab} \vec{E}^b(0) \right\rangle_T\right]_{11},
\label{delta-v}
\end{eqnarray}
where
\begin{eqnarray}
\Delta V=\frac{1}{r} \bigg( \frac{\alpha_{V_O}}{2N_c}+C_F \alpha_{V_S} \bigg) \approx \frac{N_c \alpha_s}{2r}
\end{eqnarray}
and is the potential difference between the singlet ground state and the
octet state excited by the color electric field. $d$ is the number of
the dimension in the regularization of the momentum integral.
In Ref. \cite{Brambilla08}, the electric propagator is calculated in
thermal perturbation in various limits.

To obtain the formula for the  2nd order Stark effect Eq.~\eqref{delta-v}, we have to take the
following steps.

\begin{enumerate}
 \item First, to extract the contribution from the lowest dimensional operator, we take,
       \begin{equation}
	\left[\left\langle \vec{E}^a(t) \phi(t,0)_{ab} \vec{E}^b(0) \right\rangle_T\right]_{11} \rightarrow
	 \left[\left\langle \vec{E}^a(0)  \vec{E}^a(0) \right\rangle_T\right]_{11}.
       \end{equation}
       Moreover, since we are interested in temperatures near $T_c$, instead of
       using thermal perturbation to calculate the temperature dependent
       part of the electric condensate, we use the nonperturbative value
       extracted from lattice QCD.  This approximation is valid as long
       as the scales in the matrix element is smaller than the separation
       scale.
 \item We take the matrix element of Eq.~\eqref{delta-v} for the ground
       state charmonium, and calculate it using the relative momentum
       between the $c \bar{c}$ quarks.

       \begin{equation}
	\delta E_{J/\psi}= -ig^2 \frac{T_F}{N_c} \frac{-i}{d-1} \int
	 \frac{d^3p}{(2 \pi)^3} \frac{r^2}{E_O-E_{J/\psi}} | \psi(p)|^2 \langle
	 E^a(0)E^a(0)\rangle_T
	 \label{stark-2}
       \end{equation}

 \item We assume that the energies for the intermediate octet charmonium
       sate and the initial ground state can be written as follows,

       \begin{align}
	E_{J/\psi} & =  2m_c-\epsilon \nonumber \\
	E_O & =  2m_c+p^2/m_c,
       \end{align}
       where $\epsilon$ is the binding energy for the $J/\psi$ and $E_O$
       represents the energy of the octet continuum state.  Putting these
       energies into Eq.~\eqref{stark-2} and provided $T_F=1/2$, $d=4$
       and $N_c=3$, one obtains Eq.~\eqref{eq:stark}.
       The contribution from the repulsive coulomb potential was
       neglected in the continuum octet energy as it vanishes in the
       large $N_c$ limit taken in the Peskin formalism~\cite{peskin79}.
\end{enumerate}

\end{widetext}

The leading order OPE term taken here  is quite similar to the limit
taken in Refs.~\cite{voloshin79:_qcd,leutwyler81:_how_qcd}.  The
difference there was
the assumption of Lorentz invariance of the vacuum which is broken at
finite temperature; therefore the mass shift was proportional to the
gluon condensate.
As we have identified the approximations taken in the derivation of the
2nd order Stark effect, it would be useful to improve the formula by
taking into account the renormalization group improved potentials and
the  $1/N_c$ corrections~\cite{Brambilla05}.


\begin{thebibliography}{71}
\expandafter\ifx\csname natexlab\endcsname\relax\def\natexlab#1{#1}\fi
\expandafter\ifx\csname bibnamefont\endcsname\relax
  \def\bibnamefont#1{#1}\fi
\expandafter\ifx\csname bibfnamefont\endcsname\relax
  \def\bibfnamefont#1{#1}\fi
\expandafter\ifx\csname citenamefont\endcsname\relax
  \def\citenamefont#1{#1}\fi
\expandafter\ifx\csname url\endcsname\relax
  \def\url#1{\texttt{#1}}\fi
\expandafter\ifx\csname urlprefix\endcsname\relax\def\urlprefix{URL }\fi
\providecommand{\bibinfo}[2]{#2}
\providecommand{\eprint}[2][]{\url{#2}}

\bibitem[{\citenamefont{Matsui and Satz}(1986)}]{Matsui_PLB178}
\bibinfo{author}{\bibfnamefont{T.}~\bibnamefont{Matsui}} \bibnamefont{and}
  \bibinfo{author}{\bibfnamefont{H.}~\bibnamefont{Satz}},
  \bibinfo{journal}{Phys. Lett. B} \textbf{\bibinfo{volume}{178}},
  \bibinfo{pages}{416} (\bibinfo{year}{1986}).

\bibitem[{\citenamefont{Hashimoto et~al.}(1986)\citenamefont{Hashimoto,
  Miyamura, Hirose, and Kanki}}]{hashimoto86}
\bibinfo{author}{\bibfnamefont{T.}~\bibnamefont{Hashimoto}},
  \bibinfo{author}{\bibfnamefont{O.}~\bibnamefont{Miyamura}},
  \bibinfo{author}{\bibfnamefont{K.}~\bibnamefont{Hirose}}, \bibnamefont{and}
  \bibinfo{author}{\bibfnamefont{T.}~\bibnamefont{Kanki}},
  \bibinfo{journal}{Phys. Rev. Lett.} \textbf{\bibinfo{volume}{57}},
  \bibinfo{pages}{2123} (\bibinfo{year}{1986}).

\bibitem[{\citenamefont{Eichten et~al.}(1978)\citenamefont{Eichten, Gottfried,
  Kinoshita, Lane, and Yan}}]{eichten78:_charm}
\bibinfo{author}{\bibfnamefont{E.}~\bibnamefont{Eichten}},
  \bibinfo{author}{\bibfnamefont{K.}~\bibnamefont{Gottfried}},
  \bibinfo{author}{\bibfnamefont{T.}~\bibnamefont{Kinoshita}},
  \bibinfo{author}{\bibfnamefont{K.~D.} \bibnamefont{Lane}}, \bibnamefont{and}
  \bibinfo{author}{\bibfnamefont{T.~M.} \bibnamefont{Yan}},
  \bibinfo{journal}{Phys. Rev. D} \textbf{\bibinfo{volume}{17}},
  \bibinfo{pages}{3090} (\bibinfo{year}{1978}).

\bibitem[{\citenamefont{Asakawa and Hatsuda}(2004)}]{Asakawa_PRL92}
\bibinfo{author}{\bibfnamefont{M.}~\bibnamefont{Asakawa}} \bibnamefont{and}
  \bibinfo{author}{\bibfnamefont{T.}~\bibnamefont{Hatsuda}},
  \bibinfo{journal}{Phys. Rev. Lett.} \textbf{\bibinfo{volume}{92}},
  \bibinfo{pages}{012001} (\bibinfo{year}{2004}).

\bibitem[{\citenamefont{Datta et~al.}(2004)\citenamefont{Datta, Karsch,
  Petreczky, and Wetzorke}}]{Datta_PRD69}
\bibinfo{author}{\bibfnamefont{S.}~\bibnamefont{Datta}},
  \bibinfo{author}{\bibfnamefont{F.}~\bibnamefont{Karsch}},
  \bibinfo{author}{\bibfnamefont{P.}~\bibnamefont{Petreczky}},
  \bibnamefont{and} \bibinfo{author}{\bibfnamefont{I.}~\bibnamefont{Wetzorke}},
  \bibinfo{journal}{Phys. Rev. D} \textbf{\bibinfo{volume}{69}},
  \bibinfo{pages}{094507} (\bibinfo{year}{2004}).

\bibitem[{\citenamefont{Umeda et~al.}(2005)\citenamefont{Umeda, Nomura, and
  Matsufuru}}]{umeda05}
\bibinfo{author}{\bibfnamefont{T.}~\bibnamefont{Umeda}},
  \bibinfo{author}{\bibfnamefont{K.}~\bibnamefont{Nomura}}, \bibnamefont{and}
  \bibinfo{author}{\bibfnamefont{H.}~\bibnamefont{Matsufuru}},
  \bibinfo{journal}{Eur. Phys. J. C} \textbf{\bibinfo{volume}{39}},
  \bibinfo{pages}{9} (\bibinfo{year}{2005}).

\bibitem[{\citenamefont{Jakov\'{a}c et~al.}(2007)\citenamefont{Jakov\'{a}c,
  Petreczky, Petrov, and Velytsky}}]{jakovac07}
\bibinfo{author}{\bibfnamefont{A.}~\bibnamefont{Jakov\'{a}c}},
  \bibinfo{author}{\bibfnamefont{P.}~\bibnamefont{Petreczky}},
  \bibinfo{author}{\bibfnamefont{K.}~\bibnamefont{Petrov}}, \bibnamefont{and}
  \bibinfo{author}{\bibfnamefont{A.}~\bibnamefont{Velytsky}},
  \bibinfo{journal}{Phys. Rev. D} \textbf{\bibinfo{volume}{75}},
  \bibinfo{pages}{014506} (\bibinfo{year}{2007}).

\bibitem[{\citenamefont{Aarts et~al.}(2007)\citenamefont{Aarts, Allton, Oktay,
  Peardon, and Skullerud}}]{aarts07}
\bibinfo{author}{\bibfnamefont{G.}~\bibnamefont{Aarts}},
  \bibinfo{author}{\bibfnamefont{C.}~\bibnamefont{Allton}},
  \bibinfo{author}{\bibfnamefont{M.~B.} \bibnamefont{Oktay}},
  \bibinfo{author}{\bibfnamefont{M.}~\bibnamefont{Peardon}}, \bibnamefont{and}
  \bibinfo{author}{\bibfnamefont{J.-I.} \bibnamefont{Skullerud}},
  \bibinfo{journal}{Phys. Rev. D} \textbf{\bibinfo{volume}{76}},
  \bibinfo{pages}{094513} (\bibinfo{year}{2007}).

\bibitem[{\citenamefont{M\'{o}csy}(2009)}]{mocsy:_poten}
		For a recent review on potential models at finite
		temperature, see e.g.
\bibinfo{author}{\bibfnamefont{\'{A}.}~\bibnamefont{M\'{o}csy}},
  \bibinfo{journal}{Eur. Phys. J. C} \textbf{\bibinfo{volume}{61}},
  \bibinfo{pages}{705} (\bibinfo{year}{2009}).

\bibitem[{\citenamefont{Morita and Lee}(2008{\natexlab{a}})}]{morita_jpsiprl}
\bibinfo{author}{\bibfnamefont{K.}~\bibnamefont{Morita}} \bibnamefont{and}
  \bibinfo{author}{\bibfnamefont{S.~H.} \bibnamefont{Lee}},
  \bibinfo{journal}{Phys. Rev. Lett.} \textbf{\bibinfo{volume}{100}},
  \bibinfo{pages}{022301} (\bibinfo{year}{2008}{\natexlab{a}}).

\bibitem[{\citenamefont{Morita and Lee}(2008{\natexlab{b}})}]{morita_jpsifull}
\bibinfo{author}{\bibfnamefont{K.}~\bibnamefont{Morita}} \bibnamefont{and}
  \bibinfo{author}{\bibfnamefont{S.~H.} \bibnamefont{Lee}},
  \bibinfo{journal}{Phys. Rev. C} \textbf{\bibinfo{volume}{77}},
  \bibinfo{pages}{064904} (\bibinfo{year}{2008}{\natexlab{b}}).

\bibitem[{\citenamefont{Song et~al.}(2009)\citenamefont{Song, Lee, and
  Morita}}]{song09}
\bibinfo{author}{\bibfnamefont{Y.}~\bibnamefont{Song}},
  \bibinfo{author}{\bibfnamefont{S.~H.} \bibnamefont{Lee}}, \bibnamefont{and}
  \bibinfo{author}{\bibfnamefont{K.}~\bibnamefont{Morita}},
  \bibinfo{journal}{Phys. Rev. C} \textbf{\bibinfo{volume}{79}},
  \bibinfo{pages}{014907} (\bibinfo{year}{2009}).

\bibitem[{\citenamefont{Lee and Morita}(2009)}]{lee_morita_stark}
\bibinfo{author}{\bibfnamefont{S.~H.} \bibnamefont{Lee}} \bibnamefont{and}
  \bibinfo{author}{\bibfnamefont{K.}~\bibnamefont{Morita}},
  \bibinfo{journal}{Phys. Rev. D} \textbf{\bibinfo{volume}{79}},
  \bibinfo{pages}{011501} (\bibinfo{year}{2009}).

\bibitem[{\citenamefont{Morita and Lee}(2010)}]{morita_borel}
\bibinfo{author}{\bibfnamefont{K.}~\bibnamefont{Morita}} \bibnamefont{and}
  \bibinfo{author}{\bibfnamefont{S.~H.} \bibnamefont{Lee}},
  \bibinfo{journal}{Phys. Rev. D} \textbf{\bibinfo{volume}{82}},
  \bibinfo{pages}{054008} (\bibinfo{year}{2010}).

\bibitem[{\citenamefont{Boyd et~al.}(1996)\citenamefont{Boyd, Engles, Karsch,
  Laermann, Legeland, L\"{u}tgemeier, and Petersson}}]{Boyd_NPB469}
\bibinfo{author}{\bibfnamefont{G.}~\bibnamefont{Boyd}},
  \bibinfo{author}{\bibfnamefont{J.}~\bibnamefont{Engles}},
  \bibinfo{author}{\bibfnamefont{F.}~\bibnamefont{Karsch}},
  \bibinfo{author}{\bibfnamefont{E.}~\bibnamefont{Laermann}},
  \bibinfo{author}{\bibfnamefont{C.}~\bibnamefont{Legeland}},
  \bibinfo{author}{\bibfnamefont{M.}~\bibnamefont{L\"{u}tgemeier}},
  \bibnamefont{and}
  \bibinfo{author}{\bibfnamefont{B.}~\bibnamefont{Petersson}},
  \bibinfo{journal}{Nucl. Phys.} \textbf{\bibinfo{volume}{B469}},
  \bibinfo{pages}{419} (\bibinfo{year}{1996}).

\bibitem[{\citenamefont{Panero}(2009)}]{panero09:_therm_qcd_n}
\bibinfo{author}{\bibfnamefont{M.}~\bibnamefont{Panero}},
  \bibinfo{journal}{Phys. Rev. Lett.} \textbf{\bibinfo{volume}{103}},
  \bibinfo{pages}{232001} (\bibinfo{year}{2009}).

\bibitem[{\citenamefont{Bors\'{a}nyi
  et~al.}(2010{\natexlab{a}})\citenamefont{Bors\'{a}nyi, Endr\"{o}di, Fodor,
  Jakov\'{a}c, Katz, Krieg, Ratti, and Szab\'{o}}}]{borsanyi10:_qcd}
\bibinfo{author}{\bibfnamefont{S.}~\bibnamefont{Bors\'{a}nyi}},
  \bibinfo{author}{\bibfnamefont{G.}~\bibnamefont{Endr\"{o}di}},
  \bibinfo{author}{\bibfnamefont{Z.}~\bibnamefont{Fodor}},
  \bibinfo{author}{\bibfnamefont{A.}~\bibnamefont{Jakov\'{a}c}},
  \bibinfo{author}{\bibfnamefont{S.~D.} \bibnamefont{Katz}},
  \bibinfo{author}{\bibfnamefont{S.}~\bibnamefont{Krieg}},
  \bibinfo{author}{\bibfnamefont{C.}~\bibnamefont{Ratti}}, \bibnamefont{and}
  \bibinfo{author}{\bibfnamefont{K.~K.} \bibnamefont{Szab\'{o}}},
  \bibinfo{journal}{JHEP} \textbf{\bibinfo{volume}{1011}}, \bibinfo{pages}{077}
  (\bibinfo{year}{2010}{\natexlab{a}}), \eprint{1007.2580}.

\bibitem[{\citenamefont{Aoki et~al.}(2006)\citenamefont{Aoki, Endr\"{o}di,
  Fodor, Katz, and Szab\'{o}}}]{aoki06}
\bibinfo{author}{\bibfnamefont{Y.}~\bibnamefont{Aoki}},
  \bibinfo{author}{\bibfnamefont{G.}~\bibnamefont{Endr\"{o}di}},
  \bibinfo{author}{\bibfnamefont{Z.}~\bibnamefont{Fodor}},
  \bibinfo{author}{\bibfnamefont{S.~D.} \bibnamefont{Katz}}, \bibnamefont{and}
  \bibinfo{author}{\bibfnamefont{K.~K.} \bibnamefont{Szab\'{o}}},
  \bibinfo{journal}{Nature} \textbf{\bibinfo{volume}{443}},
  \bibinfo{pages}{675} (\bibinfo{year}{2006}).

\bibitem[{\citenamefont{Bors\'{a}nyi
  et~al.}(2010{\natexlab{b}})\citenamefont{Bors\'{a}nyi, Fodor, Hoelbling,
  Katz, Krieg, Ratti, and Szab\'{o}}}]{borsanyi10:_in_t_qcd}
\bibinfo{author}{\bibfnamefont{S.}~\bibnamefont{Bors\'{a}nyi}},
  \bibinfo{author}{\bibfnamefont{Z.}~\bibnamefont{Fodor}},
  \bibinfo{author}{\bibfnamefont{C.}~\bibnamefont{Hoelbling}},
  \bibinfo{author}{\bibfnamefont{S.~D.} \bibnamefont{Katz}},
  \bibinfo{author}{\bibfnamefont{S.}~\bibnamefont{Krieg}},
  \bibinfo{author}{\bibfnamefont{C.}~\bibnamefont{Ratti}}, \bibnamefont{and}
  \bibinfo{author}{\bibfnamefont{K.~K.} \bibnamefont{Szab\'{o}}},
  \bibinfo{journal}{JHEP} \textbf{\bibinfo{volume}{1009}}, \bibinfo{pages}{073}
  (\bibinfo{year}{2010}{\natexlab{b}}).

\bibitem[{\citenamefont{Cheng et~al.}(2008)\citenamefont{Cheng, Christ, Datta,
  van~der Heide, Jung, Karsch, Kaczmarek, Laermann, Mawhinney, Miao
  et~al.}}]{cheng08}
\bibinfo{author}{\bibfnamefont{M.}~\bibnamefont{Cheng}},
  \bibinfo{author}{\bibfnamefont{N.~H.} \bibnamefont{Christ}},
  \bibinfo{author}{\bibfnamefont{S.}~\bibnamefont{Datta}},
  \bibinfo{author}{\bibfnamefont{J.}~\bibnamefont{van~der Heide}},
  \bibinfo{author}{\bibfnamefont{C.}~\bibnamefont{Jung}},
  \bibinfo{author}{\bibfnamefont{F.}~\bibnamefont{Karsch}},
  \bibinfo{author}{\bibfnamefont{O.}~\bibnamefont{Kaczmarek}},
  \bibinfo{author}{\bibfnamefont{E.}~\bibnamefont{Laermann}},
  \bibinfo{author}{\bibfnamefont{R.~D.} \bibnamefont{Mawhinney}},
  \bibinfo{author}{\bibfnamefont{C.}~\bibnamefont{Miao}}, \bibnamefont{et~al.},
  \bibinfo{journal}{Phys. Rev. D} \textbf{\bibinfo{volume}{77}},
  \bibinfo{pages}{014511} (\bibinfo{year}{2008}).

\bibitem[{\citenamefont{Bazavov et~al.}(2009)}]{bazavov09:_equat_qcd}
\bibinfo{author}{\bibfnamefont{A.}~\bibnamefont{Bazavov}} \bibnamefont{et~al.},
  \bibinfo{journal}{Phys. Rev. D} \textbf{\bibinfo{volume}{80}},
  \bibinfo{pages}{014504} (\bibinfo{year}{2009}).

\bibitem[{\citenamefont{Hatsuda et~al.}(1993)\citenamefont{Hatsuda, Koike, and
  Lee}}]{Hatsuda93}
\bibinfo{author}{\bibfnamefont{T.}~\bibnamefont{Hatsuda}},
  \bibinfo{author}{\bibfnamefont{Y.}~\bibnamefont{Koike}}, \bibnamefont{and}
  \bibinfo{author}{\bibfnamefont{S.~H.} \bibnamefont{Lee}},
  \bibinfo{journal}{Nucl. Phys.} \textbf{\bibinfo{volume}{B394}},
  \bibinfo{pages}{221} (\bibinfo{year}{1993}).

\bibitem[{\citenamefont{Cleymans et~al.}(2006)\citenamefont{Cleymans, Oeschler,
  Redlich, and Wheaton}}]{cleymans06:_compar}
\bibinfo{author}{\bibfnamefont{J.}~\bibnamefont{Cleymans}},
  \bibinfo{author}{\bibfnamefont{H.}~\bibnamefont{Oeschler}},
  \bibinfo{author}{\bibfnamefont{K.}~\bibnamefont{Redlich}}, \bibnamefont{and}
  \bibinfo{author}{\bibfnamefont{S.}~\bibnamefont{Wheaton}},
  \bibinfo{journal}{Phys. Rev. C} \textbf{\bibinfo{volume}{73}},
  \bibinfo{pages}{034905} (\bibinfo{year}{2006}).

\bibitem[{\citenamefont{Andronic et~al.}(2006)\citenamefont{Andronic,
  Braun-Munzinger, and Stachel}}]{andronic06:_hadron}
\bibinfo{author}{\bibfnamefont{A.}~\bibnamefont{Andronic}},
  \bibinfo{author}{\bibfnamefont{P.}~\bibnamefont{Braun-Munzinger}},
  \bibnamefont{and} \bibinfo{author}{\bibfnamefont{J.}~\bibnamefont{Stachel}},
  \bibinfo{journal}{Nucl. Phys.} \textbf{\bibinfo{volume}{A772}},
  \bibinfo{pages}{167} (\bibinfo{year}{2006}).

\bibitem[{\citenamefont{Andronic et~al.}(2009)\citenamefont{Andronic, Beutler,
  Braun-Munzinger, Redlich, and Stachel}}]{andronic09:_statis}
\bibinfo{author}{\bibfnamefont{A.}~\bibnamefont{Andronic}},
  \bibinfo{author}{\bibfnamefont{F.}~\bibnamefont{Beutler}},
  \bibinfo{author}{\bibfnamefont{P.}~\bibnamefont{Braun-Munzinger}},
  \bibinfo{author}{\bibfnamefont{K.}~\bibnamefont{Redlich}}, \bibnamefont{and}
  \bibinfo{author}{\bibfnamefont{J.}~\bibnamefont{Stachel}},
  \bibinfo{journal}{Phys. Lett. B} \textbf{\bibinfo{volume}{678}},
  \bibinfo{pages}{350} (\bibinfo{year}{2009}).

\bibitem[{\citenamefont{Klingl et~al.}(1999)\citenamefont{Klingl, Kim, Lee,
  Morath, and Weise}}]{Klingl_PRL82}
\bibinfo{author}{\bibfnamefont{F.}~\bibnamefont{Klingl}},
  \bibinfo{author}{\bibfnamefont{S.}~\bibnamefont{Kim}},
  \bibinfo{author}{\bibfnamefont{S.~H.} \bibnamefont{Lee}},
  \bibinfo{author}{\bibfnamefont{P.}~\bibnamefont{Morath}}, \bibnamefont{and}
  \bibinfo{author}{\bibfnamefont{W.}~\bibnamefont{Weise}},
  \bibinfo{journal}{Phys. Rev. Lett.} \textbf{\bibinfo{volume}{82}},
  \bibinfo{pages}{3396} (\bibinfo{year}{1999}).

\bibitem[{\citenamefont{Nakamura et~al.}(2010)}]{pdg2010}
\bibinfo{author}{\bibfnamefont{K.}~\bibnamefont{Nakamura}}
  \bibnamefont{et~al.}, \bibinfo{journal}{J. Phys. G.: Nucl. Part. Phys.}
  \textbf{\bibinfo{volume}{37}}, \bibinfo{pages}{075021}
  (\bibinfo{year}{2010}).

\bibitem[{\citenamefont{Koch et~al.}(1983)\citenamefont{Koch, Rafelski, and
  Greiner}}]{koch83:_stran}
\bibinfo{author}{\bibfnamefont{P.}~\bibnamefont{Koch}},
  \bibinfo{author}{\bibfnamefont{J.}~\bibnamefont{Rafelski}}, \bibnamefont{and}
  \bibinfo{author}{\bibfnamefont{W.}~\bibnamefont{Greiner}},
  \bibinfo{journal}{Phys. Lett.} \textbf{\bibinfo{volume}{123B}},
  \bibinfo{pages}{151} (\bibinfo{year}{1983}).

\bibitem[{\citenamefont{D\"{u}rr et~al.}(2008)\citenamefont{D\"{u}rr, Fodor,
  Frison, Hoelbling, Hoffmann, Katz, Krieg, Kurth, Lellouch, Lippert
  et~al.}}]{Durr08:_science}
\bibinfo{author}{\bibfnamefont{S.}~\bibnamefont{D\"{u}rr}},
  \bibinfo{author}{\bibfnamefont{Z.}~\bibnamefont{Fodor}},
  \bibinfo{author}{\bibfnamefont{J.}~\bibnamefont{Frison}},
  \bibinfo{author}{\bibfnamefont{C.}~\bibnamefont{Hoelbling}},
  \bibinfo{author}{\bibfnamefont{R.}~\bibnamefont{Hoffmann}},
  \bibinfo{author}{\bibfnamefont{S.~D.} \bibnamefont{Katz}},
  \bibinfo{author}{\bibfnamefont{S.}~\bibnamefont{Krieg}},
  \bibinfo{author}{\bibfnamefont{T.}~\bibnamefont{Kurth}},
  \bibinfo{author}{\bibfnamefont{L.}~\bibnamefont{Lellouch}},
  \bibinfo{author}{\bibfnamefont{T.}~\bibnamefont{Lippert}},
  \bibnamefont{et~al.}, \bibinfo{journal}{Science}
  \textbf{\bibinfo{volume}{322}}, \bibinfo{pages}{1224} (\bibinfo{year}{2008}).

\bibitem[{\citenamefont{Borasoy and Mei{\ss}ner}(1996)}]{borasoy96:_baryon}
\bibinfo{author}{\bibfnamefont{B.}~\bibnamefont{Borasoy}} \bibnamefont{and}
  \bibinfo{author}{\bibfnamefont{U.~G.} \bibnamefont{Mei{\ss}ner}},
  \bibinfo{journal}{Phys. Lett. B} \textbf{\bibinfo{volume}{365}},
  \bibinfo{pages}{285} (\bibinfo{year}{1996}).

\bibitem[{\citenamefont{Gl\"{u}ck et~al.}(1992)\citenamefont{Gl\"{u}ck, Reya,
  and Vogt}}]{gluck92:_parton}
\bibinfo{author}{\bibfnamefont{M.}~\bibnamefont{Gl\"{u}ck}},
  \bibinfo{author}{\bibfnamefont{E.}~\bibnamefont{Reya}}, \bibnamefont{and}
  \bibinfo{author}{\bibfnamefont{A.}~\bibnamefont{Vogt}}, \bibinfo{journal}{Z.
  Phys. C} \textbf{\bibinfo{volume}{53}}, \bibinfo{pages}{127}
  (\bibinfo{year}{1992}).

\bibitem[{\citenamefont{Gl\"{u}ck et~al.}(1999)\citenamefont{Gl\"{u}ck, Reya,
  and Schienbein}}]{gluck99:_pionic}
\bibinfo{author}{\bibfnamefont{M.}~\bibnamefont{Gl\"{u}ck}},
  \bibinfo{author}{\bibfnamefont{E.}~\bibnamefont{Reya}}, \bibnamefont{and}
  \bibinfo{author}{\bibfnamefont{I.}~\bibnamefont{Schienbein}},
  \bibinfo{journal}{Eur. Phys. J. C} \textbf{\bibinfo{volume}{10}},
  \bibinfo{pages}{313} (\bibinfo{year}{1999}).

\bibitem[{\citenamefont{Bazavov and Petreczky}(2010)}]{bazavov10:_taste_qcd}
\bibinfo{author}{\bibfnamefont{A.}~\bibnamefont{Bazavov}} \bibnamefont{and}
  \bibinfo{author}{\bibfnamefont{P.}~\bibnamefont{Petreczky}},
  \bibinfo{journal}{PoS (LATTICE2010)} p. \bibinfo{pages}{169}
  (\bibinfo{year}{2010}).

\bibitem[{\citenamefont{Majumder and
  M\"{u}ller}(2010)}]{majumder10:_hadron_qcd}
\bibinfo{author}{\bibfnamefont{A.}~\bibnamefont{Majumder}} \bibnamefont{and}
  \bibinfo{author}{\bibfnamefont{B.}~\bibnamefont{M\"{u}ller}},
  \bibinfo{journal}{Phys. Rev. Lett.} \textbf{\bibinfo{volume}{105}},
  \bibinfo{pages}{252002} (\bibinfo{year}{2010}).

\bibitem[{\citenamefont{Rischke et~al.}(1991)\citenamefont{Rischke, Gorenstein,
  St\"{o}cker, and Greiner}}]{rischke91}
\bibinfo{author}{\bibfnamefont{D.~H.} \bibnamefont{Rischke}},
  \bibinfo{author}{\bibfnamefont{M.~I.} \bibnamefont{Gorenstein}},
  \bibinfo{author}{\bibfnamefont{H.}~\bibnamefont{St\"{o}cker}},
  \bibnamefont{and} \bibinfo{author}{\bibfnamefont{W.}~\bibnamefont{Greiner}},
  \bibinfo{journal}{Z. Phys. C} \textbf{\bibinfo{volume}{51}},
  \bibinfo{pages}{485} (\bibinfo{year}{1991}).

\bibitem[{\citenamefont{Aggarwal
  et~al.}(2011)}]{aggarwal11:_scalin_proper_at_freez_out}
\bibinfo{author}{\bibfnamefont{M.~M.} \bibnamefont{Aggarwal}}
  \bibnamefont{et~al.} (\bibinfo{collaboration}{STAR Collaboration}),
  \bibinfo{journal}{Phys. Rev. C} \textbf{\bibinfo{volume}{83}},
  \bibinfo{pages}{034910} (\bibinfo{year}{2011}).

\bibitem[{\citenamefont{Kaczmarek et~al.}(2011)\citenamefont{Kaczmarek, Karsch,
  Laermann, Miao, Mukherjee, Petreczky, Schmidt, Soeldner, and
  Unger}}]{kaczmarek11:_phase_qcd}
\bibinfo{author}{\bibfnamefont{O.}~\bibnamefont{Kaczmarek}},
  \bibinfo{author}{\bibfnamefont{F.}~\bibnamefont{Karsch}},
  \bibinfo{author}{\bibfnamefont{E.}~\bibnamefont{Laermann}},
  \bibinfo{author}{\bibfnamefont{C.}~\bibnamefont{Miao}},
  \bibinfo{author}{\bibfnamefont{S.}~\bibnamefont{Mukherjee}},
  \bibinfo{author}{\bibfnamefont{P.}~\bibnamefont{Petreczky}},
  \bibinfo{author}{\bibfnamefont{C.}~\bibnamefont{Schmidt}},
  \bibinfo{author}{\bibfnamefont{W.}~\bibnamefont{Soeldner}}, \bibnamefont{and}
  \bibinfo{author}{\bibfnamefont{W.}~\bibnamefont{Unger}},
  \bibinfo{journal}{Phys. Rev. D} \textbf{\bibinfo{volume}{83}},
  \bibinfo{pages}{014504} (\bibinfo{year}{2011}).

\bibitem[{\citenamefont{Peskin}(1979)}]{peskin79}
\bibinfo{author}{\bibfnamefont{M.~E.} \bibnamefont{Peskin}},
  \bibinfo{journal}{Nucl. Phys. B} \textbf{\bibinfo{volume}{156}},
  \bibinfo{pages}{365} (\bibinfo{year}{1979}).

\bibitem[{\citenamefont{Luke et~al.}(1992)\citenamefont{Luke, Manohar, and
  Savage}}]{luke92}
\bibinfo{author}{\bibfnamefont{M.}~\bibnamefont{Luke}},
  \bibinfo{author}{\bibfnamefont{A.~V.} \bibnamefont{Manohar}},
  \bibnamefont{and} \bibinfo{author}{\bibfnamefont{M.~J.}
  \bibnamefont{Savage}}, \bibinfo{journal}{Phys. Lett. B}
  \textbf{\bibinfo{volume}{288}}, \bibinfo{pages}{355} (\bibinfo{year}{1992}).

\bibitem[{\citenamefont{Lee and Ko}(2003)}]{lee03}
\bibinfo{author}{\bibfnamefont{S.~H.} \bibnamefont{Lee}} \bibnamefont{and}
  \bibinfo{author}{\bibfnamefont{C.~M.} \bibnamefont{Ko}},
  \bibinfo{journal}{Phys. Rev. C} \textbf{\bibinfo{volume}{67}},
  \bibinfo{pages}{038202} (\bibinfo{year}{2003}).

\bibitem[{\citenamefont{Andronic et~al.}(2008)\citenamefont{Andronic,
  Braun-Munzinger, Redlich, and Stachel}}]{andronic08}
\bibinfo{author}{\bibfnamefont{A.}~\bibnamefont{Andronic}},
  \bibinfo{author}{\bibfnamefont{P.}~\bibnamefont{Braun-Munzinger}},
  \bibinfo{author}{\bibfnamefont{K.}~\bibnamefont{Redlich}}, \bibnamefont{and}
  \bibinfo{author}{\bibfnamefont{J.}~\bibnamefont{Stachel}},
  \bibinfo{journal}{Phys. Lett. B} \textbf{\bibinfo{volume}{659}},
  \bibinfo{pages}{149} (\bibinfo{year}{2008}).

\bibitem[{\citenamefont{Bertlmann}(1982)}]{bertlmann82}
\bibinfo{author}{\bibfnamefont{R.~A.} \bibnamefont{Bertlmann}},
  \bibinfo{journal}{Nucl. Phys.} \textbf{\bibinfo{volume}{B204}},
  \bibinfo{pages}{387} (\bibinfo{year}{1982}).

\bibitem[{\citenamefont{Furnstahl et~al.}(1990)\citenamefont{Furnstahl,
  Hatsuda, and Lee}}]{Furnstahl_PRD42}
\bibinfo{author}{\bibfnamefont{R.~J.} \bibnamefont{Furnstahl}},
  \bibinfo{author}{\bibfnamefont{T.}~\bibnamefont{Hatsuda}}, \bibnamefont{and}
  \bibinfo{author}{\bibfnamefont{S.~H.} \bibnamefont{Lee}},
  \bibinfo{journal}{Phys. Rev. D} \textbf{\bibinfo{volume}{42}},
  \bibinfo{pages}{1744} (\bibinfo{year}{1990}).

\bibitem[{\citenamefont{Kaczmarek and Zantow}(2005)}]{kaczmarek05:_static_qcd}
\bibinfo{author}{\bibfnamefont{O.}~\bibnamefont{Kaczmarek}} \bibnamefont{and}
  \bibinfo{author}{\bibfnamefont{F.}~\bibnamefont{Zantow}},
  \bibinfo{journal}{Phys. Rev. D} \textbf{\bibinfo{volume}{71}},
  \bibinfo{pages}{114510} (\bibinfo{year}{2005}).

\bibitem[{\citenamefont{Reinders et~al.}(1985)\citenamefont{Reinders,
  Rubinstein, and Yazaki}}]{reinders85}
\bibinfo{author}{\bibfnamefont{L.~J.} \bibnamefont{Reinders}},
  \bibinfo{author}{\bibfnamefont{H.}~\bibnamefont{Rubinstein}},
  \bibnamefont{and} \bibinfo{author}{\bibfnamefont{S.}~\bibnamefont{Yazaki}},
  \bibinfo{journal}{Phys.~Rept.} \textbf{\bibinfo{volume}{127}},
  \bibinfo{pages}{1} (\bibinfo{year}{1985}).

\bibitem[{\citenamefont{Ioffe}(2006)}]{ioffe06:_qcd_at_low_energ}
\bibinfo{author}{\bibfnamefont{B.~L.} \bibnamefont{Ioffe}},
  \bibinfo{journal}{Prog. Part. Nucl. Phys.} \textbf{\bibinfo{volume}{56}},
  \bibinfo{pages}{232} (\bibinfo{year}{2006}).

\bibitem[{\citenamefont{Shifman et~al.}(1979)\citenamefont{Shifman, Vainshtein,
  and Zakharov}}]{Shifman_NPB147}
\bibinfo{author}{\bibfnamefont{M.~A.} \bibnamefont{Shifman}},
  \bibinfo{author}{\bibfnamefont{A.~I.} \bibnamefont{Vainshtein}},
  \bibnamefont{and} \bibinfo{author}{\bibfnamefont{V.~I.}
  \bibnamefont{Zakharov}}, \bibinfo{journal}{Nucl. Phys.}
  \textbf{\bibinfo{volume}{B147}}, \bibinfo{pages}{385} (\bibinfo{year}{1979}).

\bibitem[{\citenamefont{Bochekarev and Shaposhnikov}(1986)}]{Bochekarev86}
\bibinfo{author}{\bibfnamefont{A.~I.} \bibnamefont{Bochekarev}}
  \bibnamefont{and} \bibinfo{author}{\bibfnamefont{M.~E.}
  \bibnamefont{Shaposhnikov}}, \bibinfo{journal}{Nucl. Phys.}
  \textbf{\bibinfo{volume}{B268}}, \bibinfo{pages}{220} (\bibinfo{year}{1986}).

\bibitem[{\citenamefont{Aarts and Resco}(2005)}]{aarts05}
\bibinfo{author}{\bibfnamefont{G.}~\bibnamefont{Aarts}} \bibnamefont{and}
  \bibinfo{author}{\bibfnamefont{M.~M.} \bibnamefont{Resco}},
  \bibinfo{journal}{Nucl. Phys.} \textbf{\bibinfo{volume}{B726}},
  \bibinfo{pages}{93} (\bibinfo{year}{2005}).

\bibitem[{\citenamefont{M\'{o}csy and Petreczky}(2006)}]{mocsy06}
\bibinfo{author}{\bibfnamefont{\'{A}.}~\bibnamefont{M\'{o}csy}} \bibnamefont{and}
  \bibinfo{author}{\bibfnamefont{P.}~\bibnamefont{Petreczky}},
  \bibinfo{journal}{Phys. Rev. D} \textbf{\bibinfo{volume}{73}},
  \bibinfo{pages}{074007} (\bibinfo{year}{2006}).

\bibitem[{\citenamefont{Umeda}(2007)}]{umeda07}
\bibinfo{author}{\bibfnamefont{T.}~\bibnamefont{Umeda}},
  \bibinfo{journal}{Phys. Rev. D} \textbf{\bibinfo{volume}{75}},
  \bibinfo{pages}{094502} (\bibinfo{year}{2007}).

\bibitem[{\citenamefont{Koike}(1995)}]{koike95:_qcd}
\bibinfo{author}{\bibfnamefont{Y.}~\bibnamefont{Koike}},
  \bibinfo{journal}{Phys. Rev. C} \textbf{\bibinfo{volume}{51}},
  \bibinfo{pages}{1488} (\bibinfo{year}{1995}).

\bibitem[{\citenamefont{Hatsuda et~al.}(1995)\citenamefont{Hatsuda, Lee, and
  Shiomi}}]{hatsuda95:_qcd}
\bibinfo{author}{\bibfnamefont{T.}~\bibnamefont{Hatsuda}},
  \bibinfo{author}{\bibfnamefont{S.~H.} \bibnamefont{Lee}}, \bibnamefont{and}
  \bibinfo{author}{\bibfnamefont{H.}~\bibnamefont{Shiomi}},
  \bibinfo{journal}{Phys. Rev. C} \textbf{\bibinfo{volume}{52}},
  \bibinfo{pages}{3364} (\bibinfo{year}{1995}).

\bibitem[{\citenamefont{Kim and Lee}(2001)}]{Kim01}
\bibinfo{author}{\bibfnamefont{S.}~\bibnamefont{Kim}} \bibnamefont{and}
  \bibinfo{author}{\bibfnamefont{S.~H.} \bibnamefont{Lee}},
  \bibinfo{journal}{Nucl. Phys.} \textbf{\bibinfo{volume}{A679}},
  \bibinfo{pages}{517} (\bibinfo{year}{2001}).

\bibitem[{\citenamefont{Rapp et~al.}(2010)\citenamefont{Rapp, Blaschke, and
  Crochet}}]{rapp:_charm}
\bibinfo{author}{\bibfnamefont{R.}~\bibnamefont{Rapp}},
  \bibinfo{author}{\bibfnamefont{D.}~\bibnamefont{Blaschke}}, \bibnamefont{and}
  \bibinfo{author}{\bibfnamefont{P.}~\bibnamefont{Crochet}},
  \bibinfo{journal}{Prog. Part. Nucl. Phys.} \textbf{\bibinfo{volume}{65}},
  \bibinfo{pages}{209} (\bibinfo{year}{2010}).

\bibitem[{\citenamefont{Zhao and Rapp}(2008)}]{zhao08:_trans_momen_spect_of_j}
\bibinfo{author}{\bibfnamefont{X.}~\bibnamefont{Zhao}} \bibnamefont{and}
  \bibinfo{author}{\bibfnamefont{R.}~\bibnamefont{Rapp}},
  \bibinfo{journal}{Phys. Lett. B} \textbf{\bibinfo{volume}{664}},
  \bibinfo{pages}{253} (\bibinfo{year}{2008}).

\bibitem[{\citenamefont{M\'{o}csy and Petreczky}(2008)}]{mocsy08}
\bibinfo{author}{\bibfnamefont{A.}~\bibnamefont{M\'{o}csy}} \bibnamefont{and}
  \bibinfo{author}{\bibfnamefont{P.}~\bibnamefont{Petreczky}},
  \bibinfo{journal}{Phys. Rev. D} \textbf{\bibinfo{volume}{77}},
  \bibinfo{pages}{014501} (\bibinfo{year}{2008}).

\bibitem[{\citenamefont{Park et~al.}(2007)\citenamefont{Park, Kim, Song, Lee,
  and Wong}}]{Park}
\bibinfo{author}{\bibfnamefont{Y.}~\bibnamefont{Park}},
  \bibinfo{author}{\bibfnamefont{K.~I.} \bibnamefont{Kim}},
  \bibinfo{author}{\bibfnamefont{T.}~\bibnamefont{Song}},
  \bibinfo{author}{\bibfnamefont{S.~H.} \bibnamefont{Lee}}, \bibnamefont{and}
  \bibinfo{author}{\bibfnamefont{C.~Y.} \bibnamefont{Wong}},
  \bibinfo{journal}{Phys. Rev. C} \textbf{\bibinfo{volume}{76}},
  \bibinfo{pages}{044907} (\bibinfo{year}{2007}).

\bibitem[{\citenamefont{Song et~al.}(2010)\citenamefont{Song, Park, and
  Lee}}]{Song:2010ix}
\bibinfo{author}{\bibfnamefont{T.}~\bibnamefont{Song}},
  \bibinfo{author}{\bibfnamefont{W.}~\bibnamefont{Park}}, \bibnamefont{and}
  \bibinfo{author}{\bibfnamefont{S.~H.} \bibnamefont{Lee}},
  \bibinfo{journal}{Phys. Rev. C} \textbf{\bibinfo{volume}{81}},
  \bibinfo{pages}{034914} (\bibinfo{year}{2010}).

\bibitem[{\citenamefont{Ga\'{z}dzicki and
  Gorenstein}(1999)}]{gazdzicki99:_eviden_j_gev}
\bibinfo{author}{\bibfnamefont{M.}~\bibnamefont{Ga\'{z}dzicki}}
  \bibnamefont{and} \bibinfo{author}{\bibfnamefont{M.~I.}
  \bibnamefont{Gorenstein}}, \bibinfo{journal}{Phys. Rev. Lett.}
  \textbf{\bibinfo{volume}{83}}, \bibinfo{pages}{4009} (\bibinfo{year}{1999}).

\bibitem[{\citenamefont{Andronic et~al.}(2003)\citenamefont{Andronic,
  Braun-Munzinger, Redlich, and Stachel}}]{andronic03:_statis_sps_rhic_lhc}
\bibinfo{author}{\bibfnamefont{A.}~\bibnamefont{Andronic}},
  \bibinfo{author}{\bibfnamefont{P.}~\bibnamefont{Braun-Munzinger}},
  \bibinfo{author}{\bibfnamefont{K.}~\bibnamefont{Redlich}}, \bibnamefont{and}
  \bibinfo{author}{\bibfnamefont{J.}~\bibnamefont{Stachel}},
  \bibinfo{journal}{Phys. Lett. B} \textbf{\bibinfo{volume}{571}},
  \bibinfo{pages}{36} (\bibinfo{year}{2003}).

\bibitem[{\citenamefont{Andronic et~al.}(2007)\citenamefont{Andronic,
  Braun-Munzinger, Redlich, and Stachel}}]{andronic07_plb652}
\bibinfo{author}{\bibfnamefont{A.}~\bibnamefont{Andronic}},
  \bibinfo{author}{\bibfnamefont{P.}~\bibnamefont{Braun-Munzinger}},
  \bibinfo{author}{\bibfnamefont{K.}~\bibnamefont{Redlich}}, \bibnamefont{and}
  \bibinfo{author}{\bibfnamefont{J.}~\bibnamefont{Stachel}},
  \bibinfo{journal}{Phys. Lett. B} \textbf{\bibinfo{volume}{652}},
  \bibinfo{pages}{259} (\bibinfo{year}{2007}).

\bibitem[{\citenamefont{Karsch et~al.}(2006)\citenamefont{Karsch, Kharzeev, and
  Satz}}]{Karsch06}
\bibinfo{author}{\bibfnamefont{F.}~\bibnamefont{Karsch}},
  \bibinfo{author}{\bibfnamefont{D.}~\bibnamefont{Kharzeev}}, \bibnamefont{and}
  \bibinfo{author}{\bibfnamefont{H.}~\bibnamefont{Satz}},
  \bibinfo{journal}{Phys. Lett. B} \textbf{\bibinfo{volume}{637}},
  \bibinfo{pages}{75} (\bibinfo{year}{2006}).

\bibitem[{\citenamefont{Alessandro et~al.}(2005)}]{alessandro05:_j_pb_pb_gev}
\bibinfo{author}{\bibfnamefont{B.}~\bibnamefont{Alessandro}}
  \bibnamefont{et~al.} (\bibinfo{collaboration}{NA50 Collaboration}),
  \bibinfo{journal}{Eur. Phys. J. C} \textbf{\bibinfo{volume}{39}},
  \bibinfo{pages}{335} (\bibinfo{year}{2005}).

\bibitem[{\citenamefont{Grandchamp and Rapp}(2002)}]{Grandchamp:2002wp}
\bibinfo{author}{\bibfnamefont{L.}~\bibnamefont{Grandchamp}} \bibnamefont{and}
  \bibinfo{author}{\bibfnamefont{R.}~\bibnamefont{Rapp}},
  \bibinfo{journal}{Nucl. Phys.} \textbf{\bibinfo{volume}{A709}},
  \bibinfo{pages}{415} (\bibinfo{year}{2002}).

\bibitem[{\citenamefont{Abt et~al.}(2009)}]{hera-b_chic}
\bibinfo{author}{\bibfnamefont{I.}~\bibnamefont{Abt}} \bibnamefont{et~al.}
  (\bibinfo{collaboration}{HERA-B Collaboration}), \bibinfo{journal}{Phys. Rev.
  D} \textbf{\bibinfo{volume}{79}}, \bibinfo{pages}{012001}
  (\bibinfo{year}{2009}).

\bibitem[{\citenamefont{Brambilla et~al.}(2000)\citenamefont{Brambilla, Pineda,
  Soto, and Vairo}}]{brambilla00:_poten_nrqcd}
\bibinfo{author}{\bibfnamefont{N.}~\bibnamefont{Brambilla}},
  \bibinfo{author}{\bibfnamefont{A.}~\bibnamefont{Pineda}},
  \bibinfo{author}{\bibfnamefont{J.}~\bibnamefont{Soto}}, \bibnamefont{and}
  \bibinfo{author}{\bibfnamefont{A.}~\bibnamefont{Vairo}},
  \bibinfo{journal}{Nucl. Phys. B} \textbf{\bibinfo{volume}{566}},
  \bibinfo{pages}{275} (\bibinfo{year}{2000}).

\bibitem[{\citenamefont{Brambilla et~al.}(2005)\citenamefont{Brambilla, Pineda,
  Soto, and Vairo}}]{Brambilla05}
\bibinfo{author}{\bibfnamefont{N.}~\bibnamefont{Brambilla}},
  \bibinfo{author}{\bibfnamefont{A.}~\bibnamefont{Pineda}},
  \bibinfo{author}{\bibfnamefont{J.}~\bibnamefont{Soto}}, \bibnamefont{and}
  \bibinfo{author}{\bibfnamefont{A.}~\bibnamefont{Vairo}},
  \bibinfo{journal}{Rev. Mod. Phys.} \textbf{\bibinfo{volume}{77}},
  \bibinfo{pages}{1423} (\bibinfo{year}{2005}).

\bibitem[{\citenamefont{Brambilla et~al.}(2008)\citenamefont{Brambilla,
  Ghiglieri, Vairo, and Petreczky}}]{Brambilla08}
\bibinfo{author}{\bibfnamefont{N.}~\bibnamefont{Brambilla}},
  \bibinfo{author}{\bibfnamefont{J.}~\bibnamefont{Ghiglieri}},
  \bibinfo{author}{\bibfnamefont{V.}~\bibnamefont{Vairo}}, \bibnamefont{and}
  \bibinfo{author}{\bibfnamefont{P.}~\bibnamefont{Petreczky}},
  \bibinfo{journal}{Phys. Rev. D} \textbf{\bibinfo{volume}{78}},
  \bibinfo{pages}{014017} (\bibinfo{year}{2008}).

\bibitem[{\citenamefont{Voloshin}(1979)}]{voloshin79:_qcd}
\bibinfo{author}{\bibfnamefont{M.~B.} \bibnamefont{Voloshin}},
  \bibinfo{journal}{Nucl. Phys. B} \textbf{\bibinfo{volume}{154}},
  \bibinfo{pages}{365} (\bibinfo{year}{1979}).

\bibitem[{\citenamefont{Leutwyler}(1981)}]{leutwyler81:_how_qcd}
\bibinfo{author}{\bibfnamefont{H.}~\bibnamefont{Leutwyler}},
  \bibinfo{journal}{Phys. Lett. B} \textbf{\bibinfo{volume}{98}},
  \bibinfo{pages}{447} (\bibinfo{year}{1981}).

\end{thebibliography}

\end{document}